\documentclass[aps,prd,showpacs,floatfix,nofootinbib,twocolumn,10pt]{revtex4}

\usepackage{color,amsmath,amssymb,graphicx,latexsym,subfigure}
\usepackage{threeparttable,txfonts}

\def\aap{Astron.\ Astrophys.\ }
\def\apj{Astrophys.\ J.\ }
\def\apjl{Astrophys.\ J.\ Lett.\ }

\def\mnras{Mon.\ Not.\ Roy.\ Astron.\ Soc.\ }

\def\prd{Phys.\ Rev.\ D\ }
\def\prl{Phys.\ Rev.\ Lett.\ }

\def\jcap{J.\ Cosmol.\ Astropart.\ Phys.\ }

\begin{document}

\title{Implications on the origin of cosmic rays in light of 10 TV 
spectral softenings}

\author{Chuan Yue$^{a}$}
\author{Peng-Xiong Ma$^{a,b}$}
\author{Qiang Yuan$^{a,b,c}$\footnote{Corresponding author: yuanq@pmo.ac.cn}}
\author{Yi-Zhong Fan$^{a,b}$\footnote{Corresponding author: yzfan@pmo.ac.cn}}
\author{Zhan-Fang Chen$^{a,b}$}
\author{Ming-Yang Cui$^a$}
\author{Hao-Ting Dai$^d$}
\author{Tie-Kuang Dong$^a$}
\author{Xiaoyuan Huang$^a$}
\author{Wei Jiang$^{a,b}$}
\author{Shi-Jun Lei$^a$}
\author{Xiang Li$^a$}
\author{Cheng-Ming Liu$^d$}
\author{Hao Liu$^a$}
\author{Yang Liu$^a$}
\author{Chuan-Ning Luo$^{a,b}$}
\author{Xu Pan$^{a,b}$}
\author{Wen-Xi Peng$^e$}
\author{Rui Qiao$^e$}
\author{Yi-Feng Wei$^d$}
\author{Li-Bo Wu$^d$}
\author{Zhi-Hui Xu$^{a,b}$}
\author{Zun-Lei Xu$^a$}
\author{Guan-Wen Yuan$^{a,b}$}
\author{Jing-Jing Zang$^a$}
\author{Ya-Peng Zhang$^f$}
\author{Yong-Jie Zhang$^f$}
\author{Yun-Long Zhang$^d$}

\affiliation{
$^a$Key Laboratory of Dark Matter and Space Astronomy, Purple Mountain
Observatory, Chinese Academy of Sciences, Nanjing 210033, China \\
$^b$School of Astronomy and Space Science, University of Science and
Technology of China, Hefei 230026, China\\
$^c$Center for High Energy Physics, Peking University, Beijing 100871, China \\
$^d$State Key Laboratory of Particle Detection and Electronics, University
of Science and Technology of China, Hefei 230026, China\\
$^e$Key Laboratory of Particle Astrophysics, Institute of High Energy
Physics, Chinese Academy of Sciences, Beijing 100049, China\\
$^f$Institute of Modern Physics, Chinese Academy of Sciences, Lanzhou
730000, China
}

\begin{abstract}

Precise measurements of the energy spectra of cosmic rays (CRs) show 
various kinds of features deviating from single power-laws, which give 
very interesting and important implications on their origin and propagation.
Previous measurements from a few balloon and space experiments
indicate the existence of spectral softenings around 10 TV for protons 
(and probably also for Helium nuclei). Very recently, the DArk Matter 
Particle Explorer (DAMPE) measurement about the proton spectrum clearly 
reveals such a softening with a high significance. Here we study the 
implications of these new measurements, as well as the groundbased 
indirect measurements, on the origin of CRs. 
We find that a single component of CRs fails to fit the spectral
softening and the air shower experiment data simultaneously.
In the framework of multiple components, we discuss two possible scenarios, 
the multiple source population scenario and the background plus nearby 
source scenario. Both scenarios give reasonable fits to the wide-band 
data from TeV to 100 PeV energies. Considering the anisotropy observations, 
the nearby source model is favored.

\end{abstract}

\date{\today}


\maketitle

\section{Introduction}

The origin of cosmic rays (CRs) remains an unresolved question after more 
than one century since their discovery. To identify the sources of CRs is 
difficult due to that the diffusive propagation of charged particles in 
the random magnetic field results in the loss of the original directions
of CRs. Precise measurements of the energy spectra of various species of
CRs are helpful in understanding their origin and propogation. The energy 
spectra of CRs from the acceleration sources are generally believed to be 
power-laws with cutoffs due to the maximum acceleration limits of specific 
types of sources. The diffusion in the Galaxy results in softenings of the 
accelerated spectra, by a power-law of $E^{-\delta}$, which reflects the 
energy-dependence of the diffusion coefficient and hence the turbulent
properties of the interstellar medium. Such an effect has been supported 
by the measurement of the secondary-to-primary flux ratios of CR nuclei 
\cite{2016PhRvL.117w1102A}.

However, several balloon and space experiments revealed remarkable
spectral hardenings of CR nuclei around a few hundred GV rigidities 
\cite{2009BRASP..73..564P,2010ApJ...714L..89A,2011Sci...332...69A,
2015PhRvL.114q1103A,2015PhRvL.115u1101A,2017PhRvL.119y1101A,
2019PhRvL.122r1102A}. These results inspire quite a number of discussions 
of their possible implications on the origin \cite{2011ApJ...729L..13O,
2011PhRvD..84d3002Y,2012ApJ...752...68V,2012APh....35..449E,
2012MNRAS.421.1209T,2013A&A...555A..48B,2017PhRvD..96b3006L}, acceleration 
\cite{2013ApJ...763...47P,2014A&A...567A..33T,2017ApJ...844L...3Z}, 
and propagation \cite{2012ApJ...752L..13T,2012PhRvL.109f1101B,
2015ApJ...803L..15T,2017PhRvD..95b3001T,2016ChPhC..40a5101J,
2016ApJ...819...54G,2018PhRvD..97f3008G,2018ApJ...869..176L} of CRs.
The AMS-02 measurements of the spectra of the secondary family of nuclei, 
Li, Be, and B, show that on average their spectra harden above $\sim200$ 
GV by $E^{0.13}$ more than that of the primary family of He, C, and O 
\cite{2018PhRvL.120b1101A}, which indicates that the spectral hardenings 
may have a propagation origin \cite{2017PhRvL.119x1101G}. Nevertheless, 
it is shown that the injection hardening scenario can also fit the data 
reasonably well in a class of propagation models with effective 
reacceleration of particles in the turbulent medium 
\cite{2018arXiv181003141Y,2018arXiv181009301N}.

Improved direct measurements of the CR spectra at higher energies 
are recently available from several experiments. Interestingly, the
CREAM \cite{2017ApJ...839....5Y} and NUCLEON \cite{2018JETPL.108....5A} 
data show hints that the CR spectra become softer for rigidities higher
than 10 TV. The precise measurement of the proton spectrum up to 100 
TeV by the Dark Matter Particle Explorer (DAMPE; 
\cite{ChangJin:550,2017APh....95....6C}) clearly reveal such a spectral
softening \cite{2019arXiv190912860A}. On the other hand, ground-based air 
shower experiemnts show that the all-particle spectrum has a so-called 
``knee'' at energies of a few PeV (e.g., \cite{2005APh....24....1A,
2007NuPhS.165...74K,2008ApJ...678.1165A,2008JPhG...35k5201G}).
Measurements of the knee of individual composition have relatively
large uncertainties \cite{2005APh....24....1A,2006PhLB..632...58T}.
A few measurements of the light composition group, e.g. proton plus 
helium nuclei, tend to suggest a knee below PeV energies
\cite{2015PhRvD..92i2005B}. Most recently, preliminary results about 
the proton plus helium spectra measured by the HAWC experiment showed
also a softening at about 30 TeV energies \cite{HAWC-ICRC2019-176}. 
Given all these progresses of the measurements, it is thus very 
interesting to investigate the implications of the wide-band direct 
and indirect measurements on the CR modeling.

\begin{figure*}[!htb]
\includegraphics[width=0.45\textwidth]{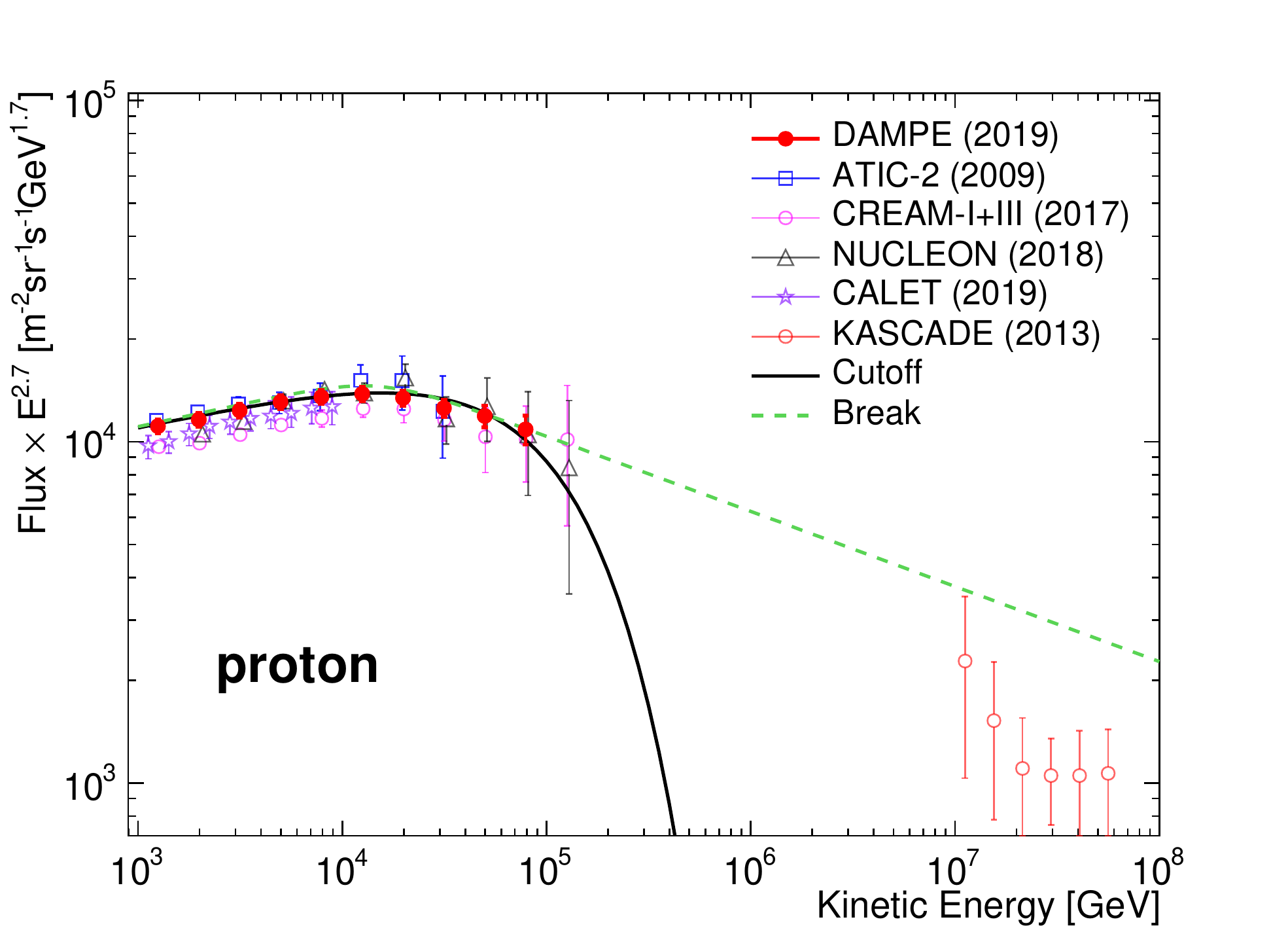}
\includegraphics[width=0.45\textwidth]{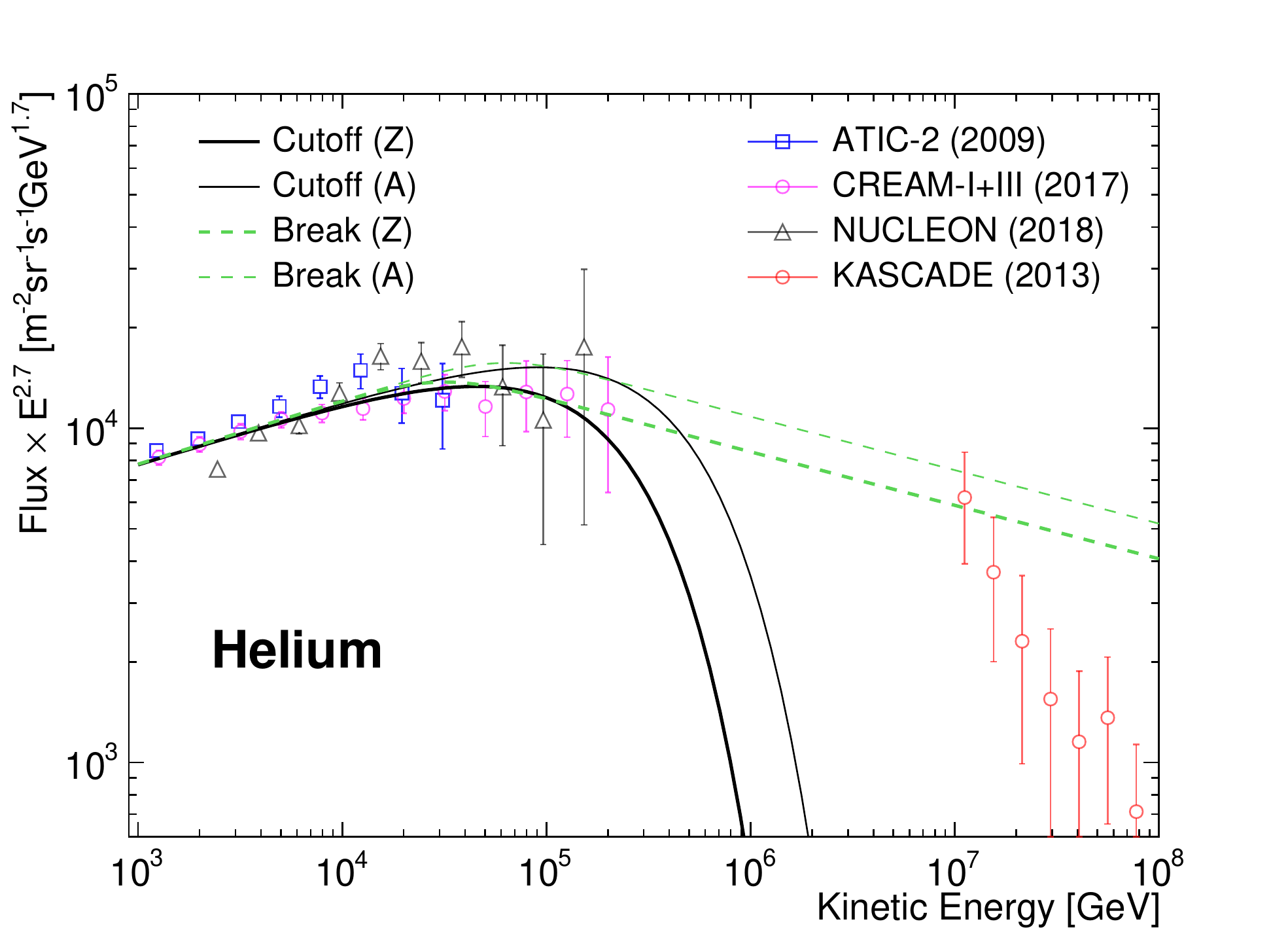}
\includegraphics[width=0.45\textwidth]{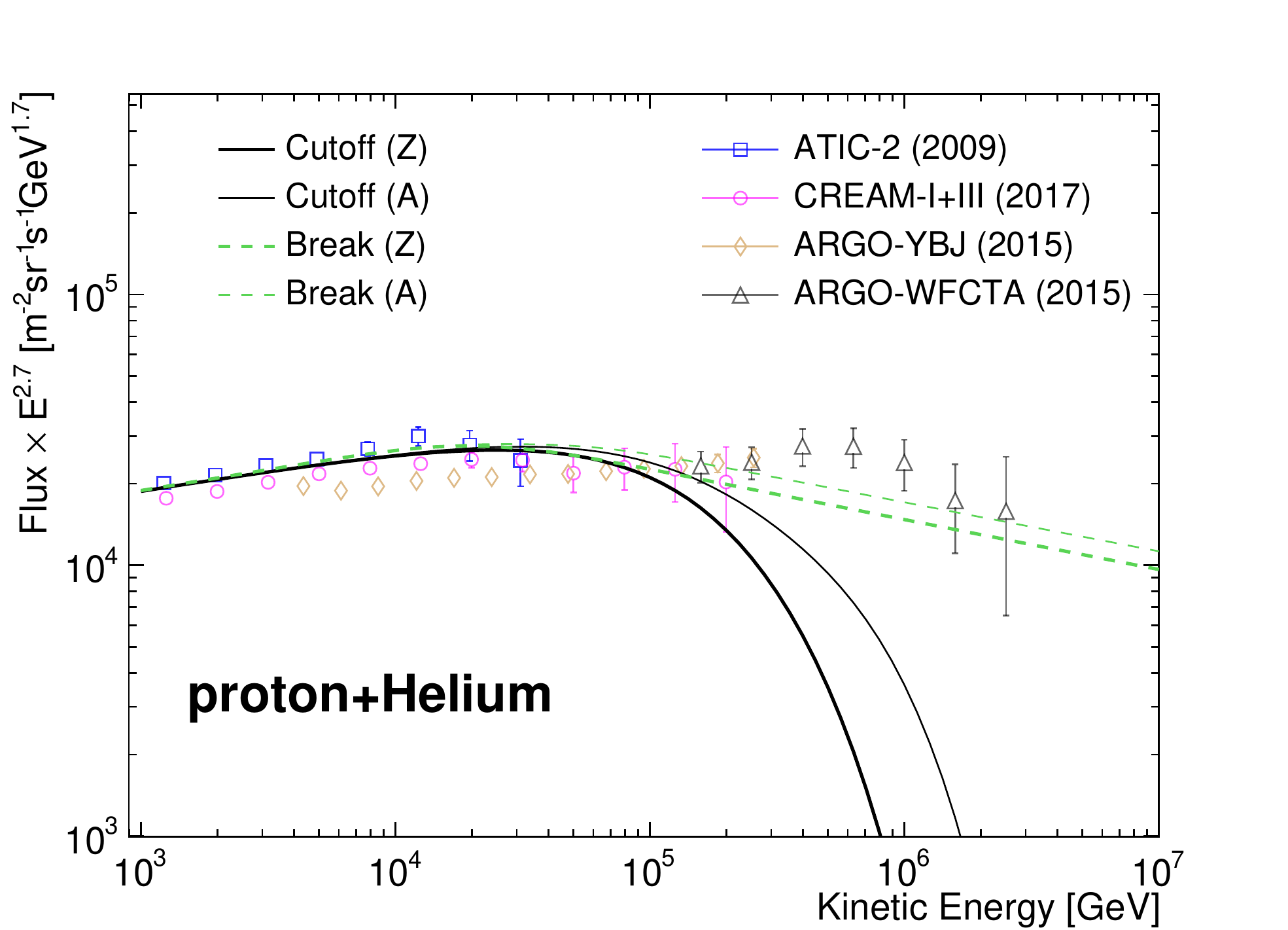}
\includegraphics[width=0.45\textwidth]{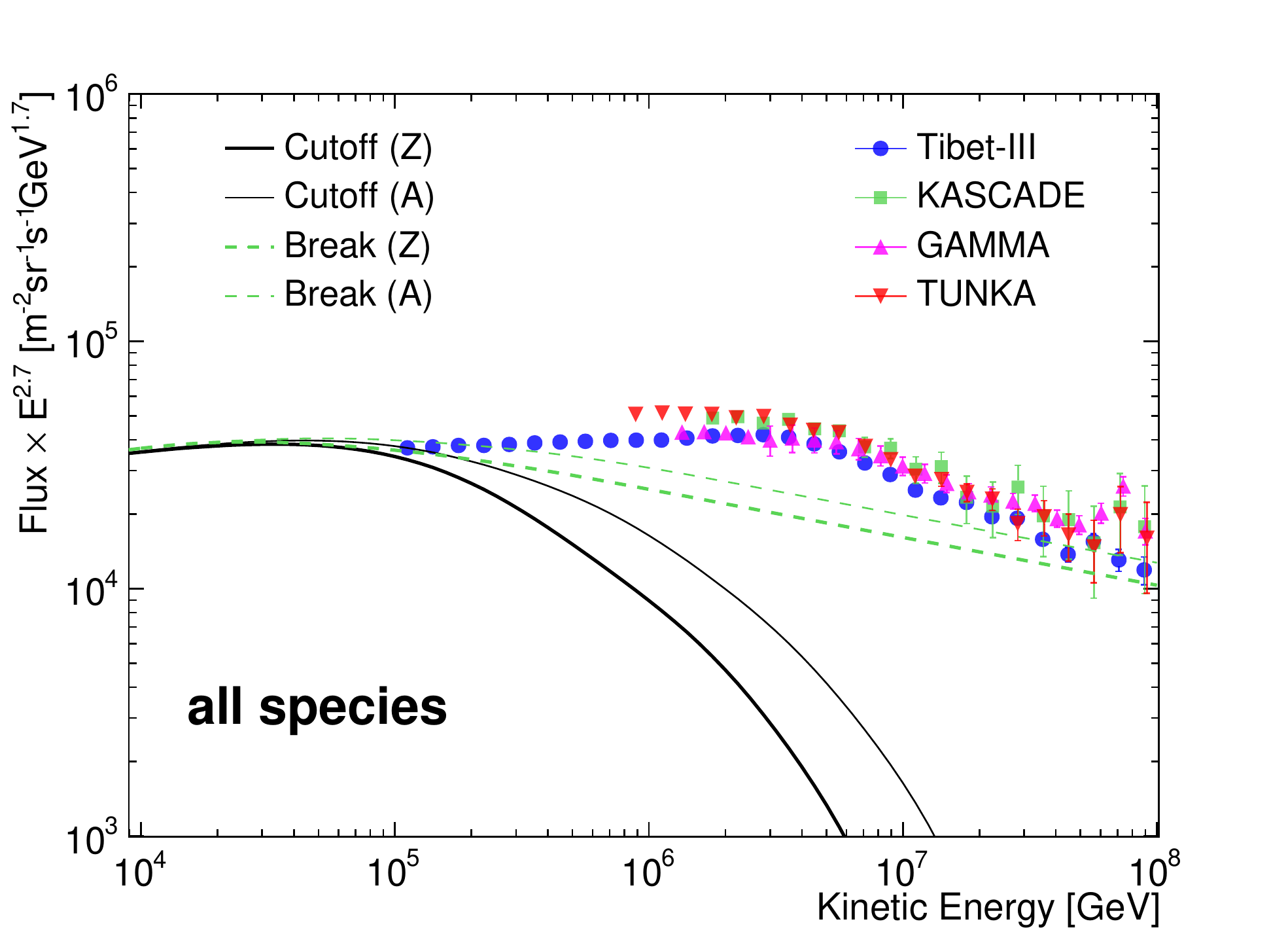}
\caption{Energy spectra of protons (top-left), Helium (top-right),
proton plus Helium (bottom-left), and all species (bottom-right).
In each panel the solid lines show the fitting results with an
exponential cutoff form (eq.(1)), and the dashed lines show the
broken power-law (eq.(2)) fitting results. The thick lines are for
the $Z$-dependent cutoff/break energies, and the thin lines are for
the $A$-dependent cases.
References of the data: protons, ATIC \cite{2009BRASP..73..564P},
CREAM \cite{2017ApJ...839....5Y}, NUCLEON \cite{2018JETPL.108....5A},
CALET \cite{2019PhRvL.122r1102A}, DAMPE \cite{2019arXiv190912860A}, 
KASCADE \cite{2013APh....47...54A}; 
Helium, ATIC \cite{2009BRASP..73..564P}, CREAM \cite{2017ApJ...839....5Y}, 
NUCLEON \cite{2018JETPL.108....5A}, KASCADE \cite{2013APh....47...54A};
p+He, ATIC \cite{2009BRASP..73..564P}, CREAM \cite{2017ApJ...839....5Y}, 
ARGO-YBJ \cite{2015PhRvD..91k2017B}, ARGO-WFCTA \cite{2015PhRvD..92i2005B}; 
all-particle, Tibet-III \cite{2008ApJ...678.1165A}, KASCADE 
\cite{2005APh....24....1A}, GAMMA \cite{2008JPhG...35k5201G}, 
TUNKA \cite{2007NuPhS.165...74K}.
\label{fig:specI}}
\end{figure*}

There are some studies based on the data available at different time 
\cite{2003APh....19..193H,2006A&A...458....1Z,2006astro.ph..7109H,
2012APh....35..801G,2013FrPhy...8..748G,2016A&A...595A..33T,
2018ChPhC..42g5103G}. In particular, several studies propose to 
account for various spectral structures using multiple populations 
of CR sources \cite{2006astro.ph..7109H,2006A&A...458....1Z,
2012APh....35..801G,2013FrPhy...8..748G}. Alternatively, if there are 
by chance one or a few nearby sources whose contributions are different 
from the sum of the other background sources, spectral structures may 
also be produced \cite{1997JPhG...23..979E,2013APh....50...33S,
2015ApJ...809L..23S,2019JCAP...10..010L,2019arXiv190100249Q,
2019arXiv190512505Q,2019arXiv190705987K}. In light of the new measurements 
of the CR spectra, in particular, by the DAMPE, we revisit the modeling 
of CR sources from TeV to 100 PeV in a phenomenological way. Our discussion 
is within the framework of the above two scenarios, i.e. multiple populations 
(denoted as model A) and nearby sources (denoted as model B), but with a 
focus on the $O(10)$ TV spectral features. Both models have good 
physical motivations. For model A, for example, the remnants of different 
types of supernovae which are smoothly distributed in the Galactic disk 
should behave differently in accelerating CR particles. The sum of their 
contributions can result in complicated spectral features. Alternatively, 
if the Earth is close to (e.g., $\lesssim 500$~pc) one single accelerator 
by chance, the distinct spectral feature from this nearby source may 
naturally give the observed spectral bumps. 
The purpose of this study is to build an overall model of CRs to describe 
as many as possible the up-to-date observational data in a wide energy range. 

\section{Origin of the spectral softening}

It is clear that the spectral softenings around $\sim10$ TV do not
correspond to the PeV knee of CRs, even for $A$-dependent knees of
various compositions. To see this explicitly, we show in 
Fig.~\ref{fig:specI} the energy spectra of protons, Helium, protons 
plus Helium, and the all-particle one, for the fitting with one single 
component of each species. We assume either an exponential cutoff 
power-law form or a broken power-law form to describe the spectral 
softenings of CR nuclei, as
\begin{equation}
\Phi_i(E)=\Phi_{0,i}\left(\frac{E}{\rm TeV}\right)^{-\gamma_i}\times
\exp\left(-\frac{E}{E_{c,i}}\right),
\end{equation}
and
\begin{equation}
\Phi_i(E)=\Phi_{0,i}\left(\frac{E}{\rm TeV}\right)^{-\gamma_i}\times
\left[1+\left(\frac{E}{E_{b,i}}\right)^s\right]^{(-\Delta\gamma/s)},
\end{equation}
where $E$ is the total energy of a particle, the subscription $i$ 
represents different nuclear species, $\gamma_i$ is the spectral index 
below the energy of the softening, $E_{b,i}$ ($E_{c,i}$) is the break 
(cutoff) energy, $s$ is a smoothness parameter, and $\Delta\gamma$ is 
the change of the spectral index above $E_{b,i}$. These parameters are 
determined through fitting to the measurements of energy spectra of 
individual species by ATIC \cite{2009BRASP..73..564P}, 
CREAM \cite{2011ApJ...728..122Y,2017ApJ...839....5Y}, NUCLEON 
\cite{2018JETPL.108....5A}, and DAMPE \cite{2019arXiv190912860A}.
For different nuclear species, we assume that the break (cutoff) energy 
$E_{b,i}$ ($E_{c,i}$) is proportional to either the atomic number $Z_i$ 
or the mass number $A_i$, i.e., $E_{b,i}=Z_i\epsilon_b$ or $A_i\epsilon_b$ 
($E_{c,i}=Z_i\epsilon_c$ or $A_i\epsilon_c$). For the broken power-law
fit, the proton spectrum suggests that $s=3.0$ and $\Delta\gamma=0.35$ 
can describe the spectral softening well. The other parameters are give 
in Table~\ref{tab:paramI}. The results show that the p+He and the 
all-particle spectra cannot be reproduced in all these fittings, and
additional spectral structures between the $O(10)$ TV softening and the
knee of CRs are expected (see also Ref.~\cite{2019arXiv191101311L}). In the 
following we discuss two natural scenarios of these spectral structures.


\begin{table*}[!htb]
\begin{center}
\caption{Spectral parameters of major CR species assuming $\sim10$~TV knees.}
\begin{tabular}{ccccccc}\hline \hline
Species & $\Phi_{0,i}$ & $\gamma_i$ & $\epsilon_b$ & $\epsilon_c$ \\ 
        & (m$^{-2}$s$^{-1}$sr$^{-1}$TeV$^{-1}$) & & (TeV) & (TeV) \\ \hline
p       & $8.79\times10^{-2}$ & 2.57 & $15$ & $120$ \\
He      & $6.20\times10^{-2}$ & 2.51 & $15$ & $120$ \\
C       & $1.05\times10^{-2}$ & 2.56 & $15$ & $120$ \\
O       & $1.35\times10^{-2}$ & 2.56 & $15$ & $120$ \\
Ne      & $4.73\times10^{-3}$ & 2.56 & $15$ & $120$ \\
Mg      & $7.43\times10^{-3}$ & 2.56 & $15$ & $120$ \\
Si      & $8.78\times10^{-3}$ & 2.56 & $15$ & $120$ \\
Fe      & $1.50\times10^{-2}$ & 2.56 & $15$ & $120$ \\
\hline \hline
\end{tabular}
\label{tab:paramI}
\end{center}
\end{table*}

\subsection{Multiple populations of CR sources}

\begin{figure*}[!htb]
\includegraphics[width=0.45\textwidth]{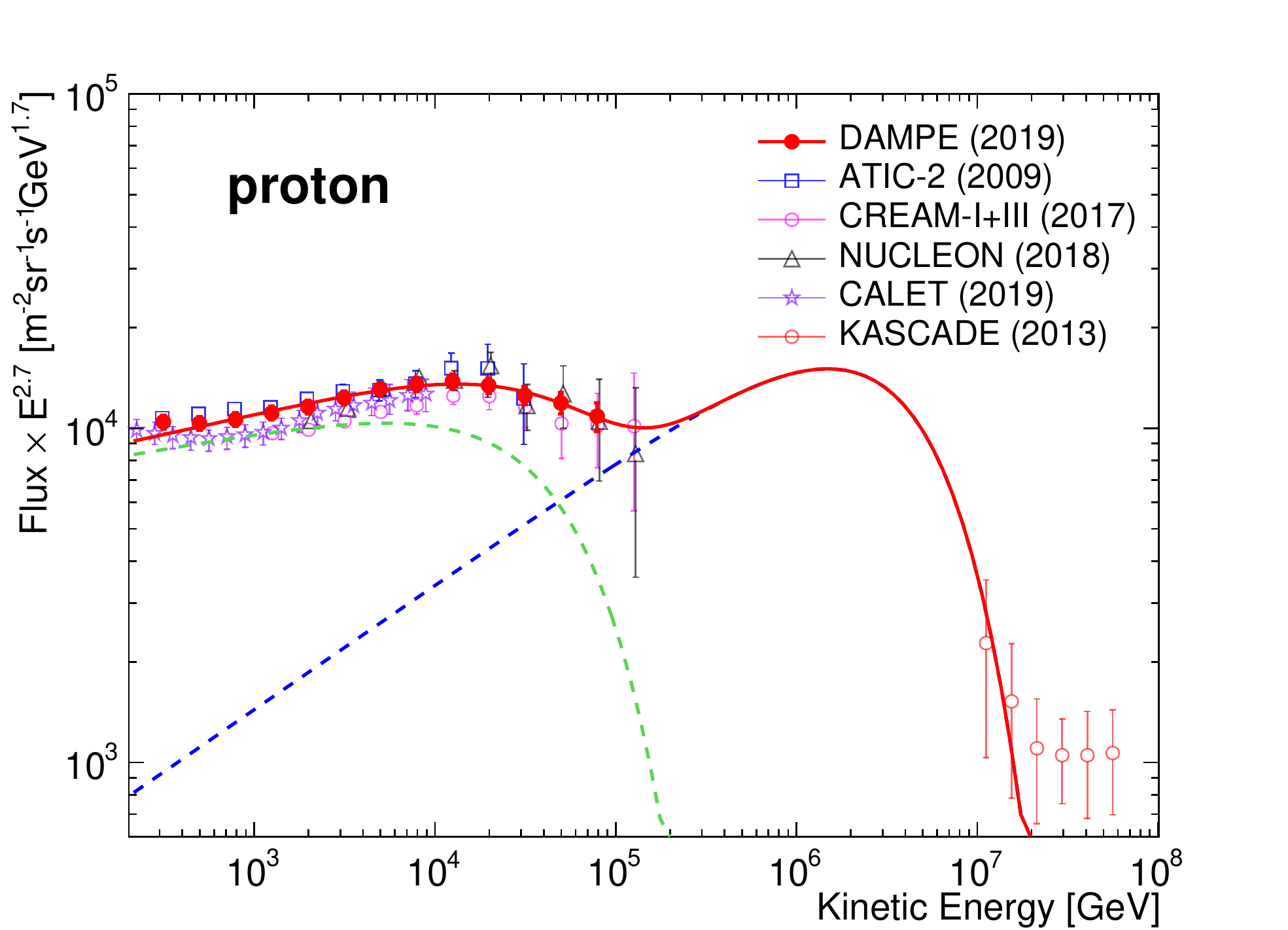}
\includegraphics[width=0.45\textwidth]{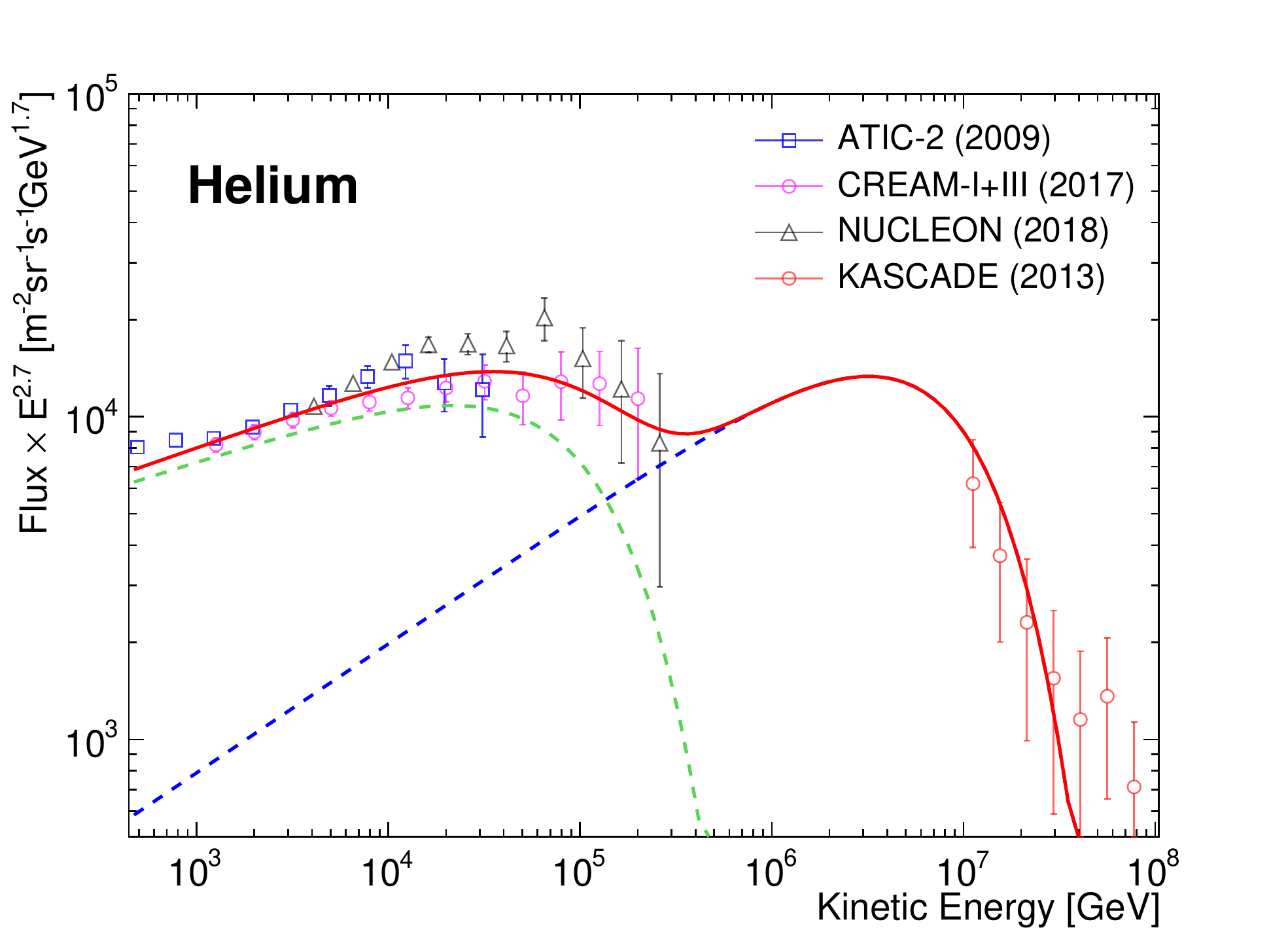}
\includegraphics[width=0.45\textwidth]{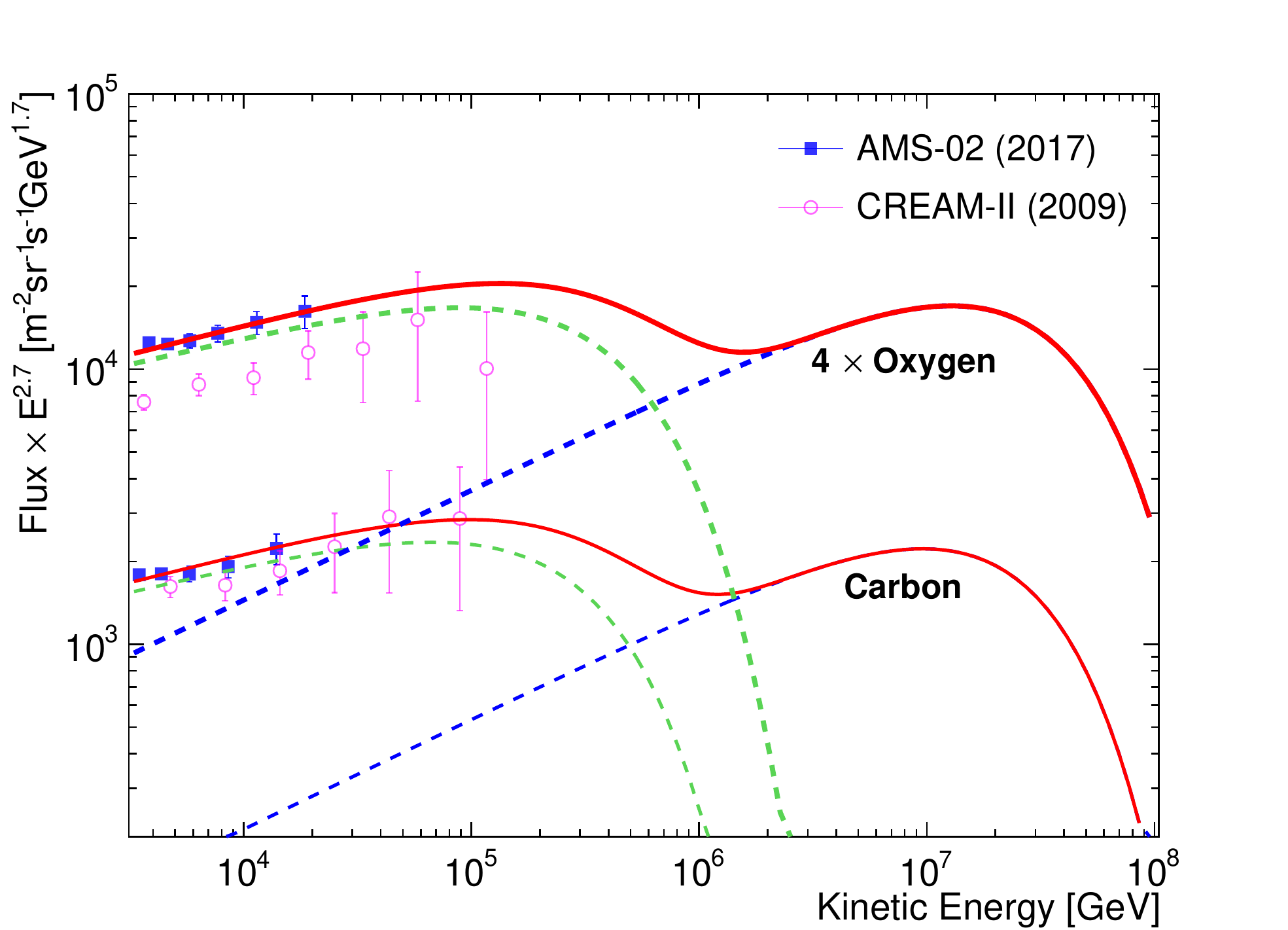}
\includegraphics[width=0.45\textwidth]{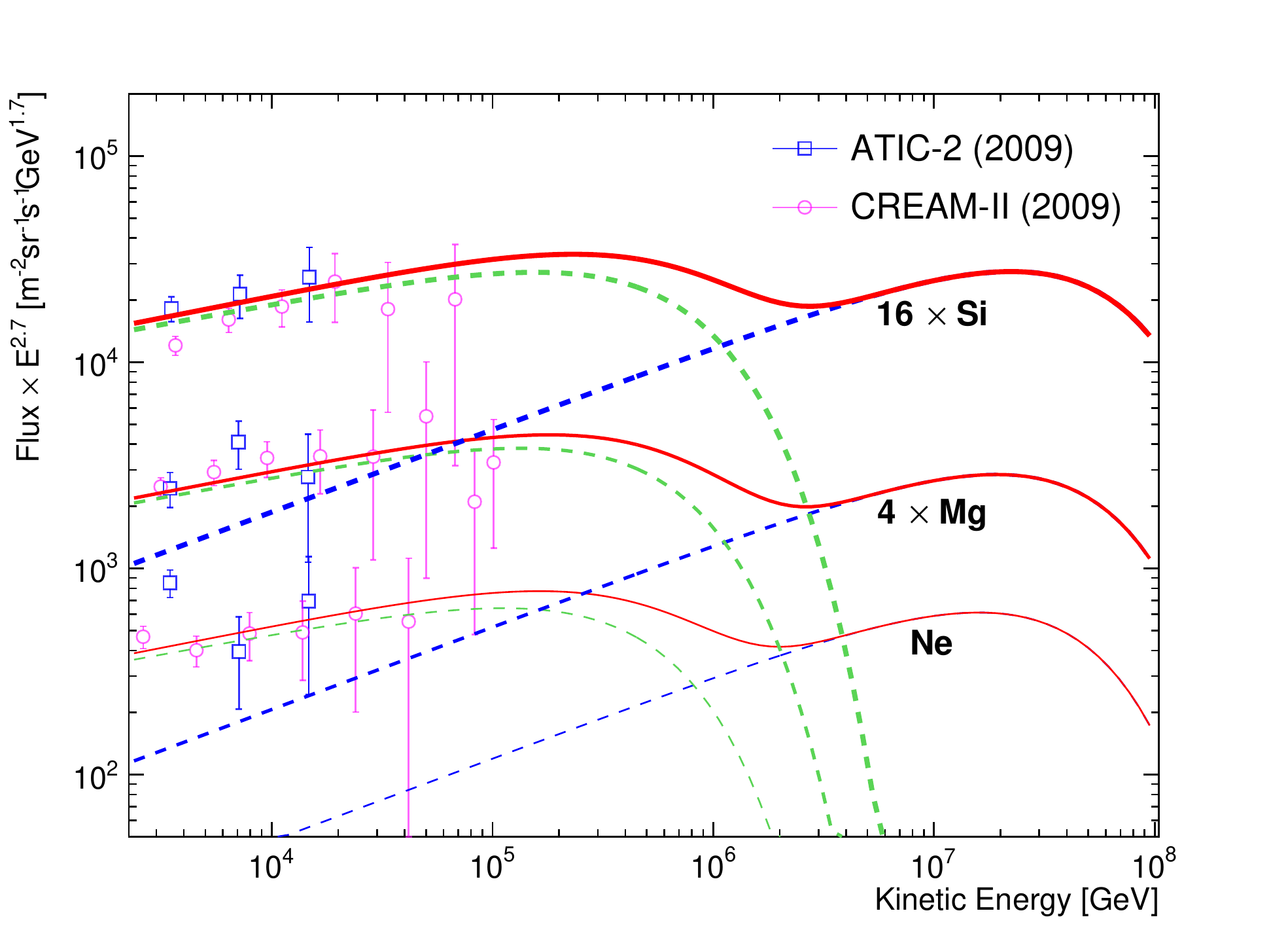}
\includegraphics[width=0.45\textwidth]{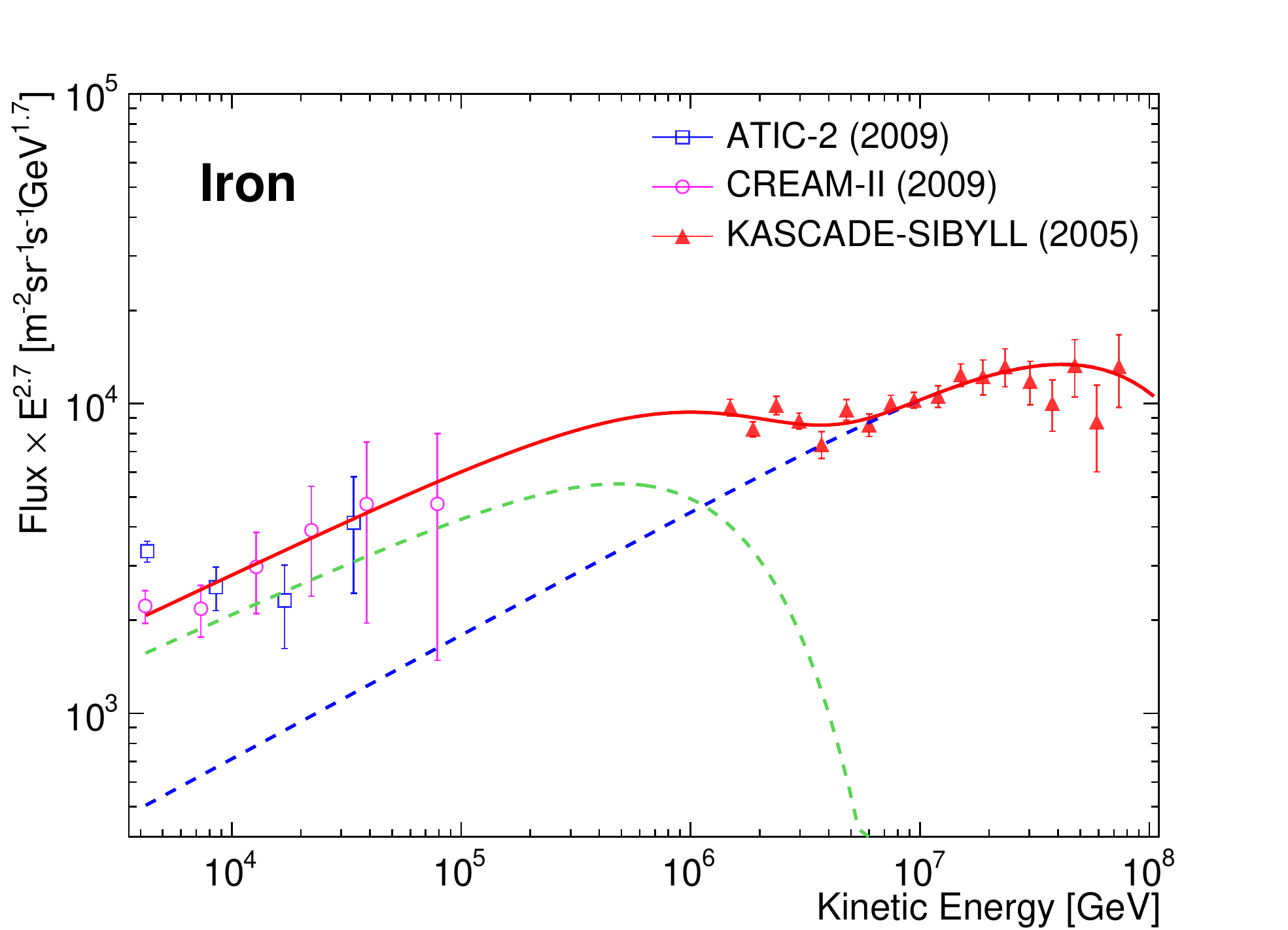}
\includegraphics[width=0.45\textwidth]{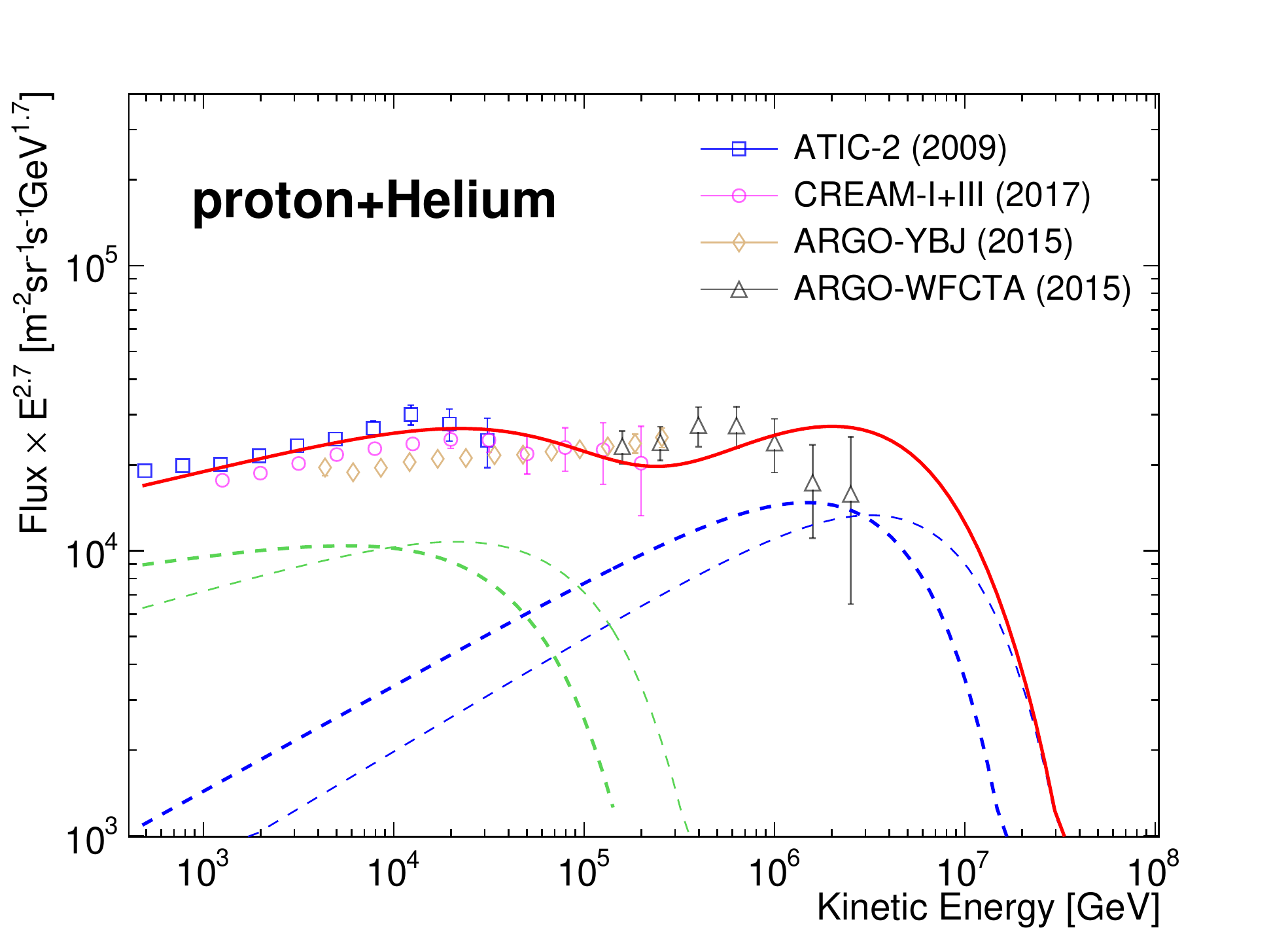}
\includegraphics[width=0.45\textwidth]{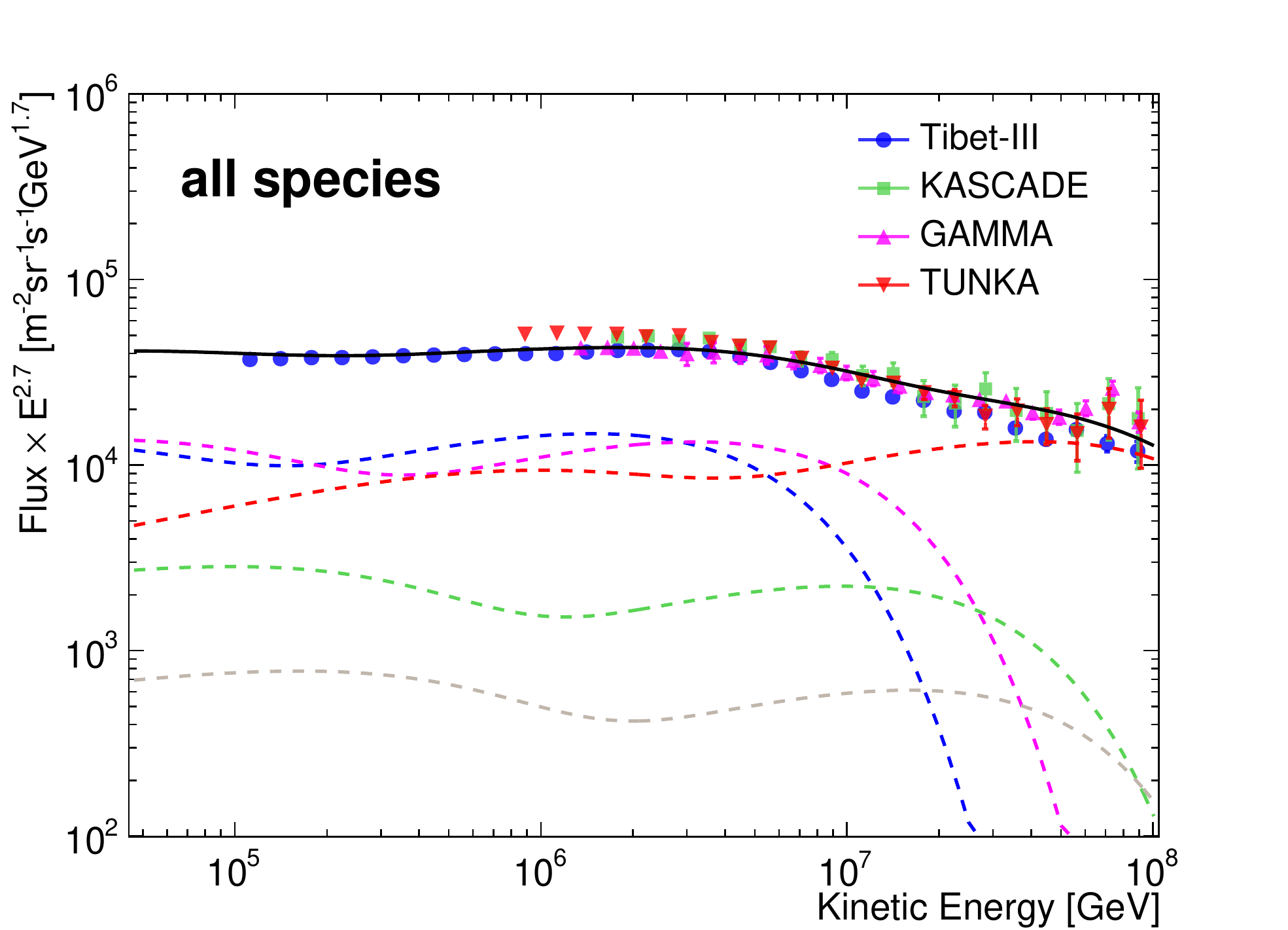}
\caption{Fitting energy spectra for model A, compared with the data.
In each panel, the green and blue dashed curves show the contributions
of each source population, and the solid curves are the total contribution.
References of the data: Carbon and Oxygen, AMS-02 \cite{2017PhRvL.119y1101A}, 
CREAM \cite{2009ApJ...707..593A}; Neon, Magnesium, and Silicon,
ATIC \cite{2009BRASP..73..564P}, CREAM \cite{2009ApJ...707..593A};
Iron, ATIC \cite{2009BRASP..73..564P}, CREAM \cite{2009ApJ...707..593A},
KASCADE \cite{2005APh....24....1A}. The other references are the same 
as in Fig.~\ref{fig:specI}.
\label{fig:specII}}
\end{figure*}

It has been widely postulated that there are more than one populations of 
CR sources in the Milky Way. For instance, supernovae of different types 
may accelerate particles to different maximum energies, giving various 
spectral features of CRs \cite{2006A&A...458....1Z,2013FrPhy...8..748G}.
Following Ref.~\cite{2013FrPhy...8..748G}, we assume that the spectrum 
of each population is described by an exponential cutoff power-law function
of eq.~(1). We further assume that the cutoff energies of different
species of each population depend on the atomic number $Z_i$, i.e., 
$E_{c,i}=Z_i\epsilon_c$. The fitting results of the major species as well 
as the all-particle spectrum are shown in Fig.~\ref{fig:specII}. 
The spectral parameters are summarized in Table \ref{tab:paramII}. 

\begin{table*}[!htb]
\begin{center}
\caption{Spectral parameters of model A.}
\begin{tabular}{cccc|ccc}\hline \hline
        & \multicolumn{3}{c|}{Pop. I} & \multicolumn{3}{c}{Pop. II} \\
Species & $\Phi_{0,i}$ & $\gamma_i$ & $\epsilon_c$ & $\Phi_{0,i}$ & $\gamma_i$ & $\epsilon_c$ \\ 
        & (m$^{-2}$s$^{-1}$sr$^{-1}$TeV$^{-1}$) & & (TeV) & (m$^{-2}$s$^{-1}$sr$^{-1}$TeV$^{-1}$) & & (TeV) \\ \hline
p       & $7.78\times10^{-2}$ & 2.60 & $56$ & $1.15\times10^{-2}$ & 2.33 & $4.0\times10^3$ \\
He      & $5.84\times10^{-2}$ & 2.51 & $56$ & $6.30\times10^{-3}$ & 2.30 & $4.0\times10^3$ \\
C       & $9.92\times10^{-3}$ & 2.50 & $56$ & $7.00\times10^{-4}$ & 2.30 & $4.0\times10^3$ \\
O       & $1.66\times10^{-2}$ & 2.50 & $56$ & $1.10\times10^{-3}$ & 2.30 & $4.0\times10^3$ \\
Ne      & $2.40\times10^{-3}$ & 2.50 & $56$ & $1.37\times10^{-4}$ & 2.30 & $4.0\times10^3$ \\
Mg      & $3.52\times10^{-3}$ & 2.50 & $56$ & $2.22\times10^{-4}$ & 2.30 & $4.0\times10^3$ \\
Si      & $6.08\times10^{-3}$ & 2.50 & $56$ & $3.71\times10^{-4}$ & 2.30 & $4.0\times10^3$ \\
Fe      & $7.78\times10^{-3}$ & 2.37 & $56$ & $2.27\times10^{-3}$ & 2.30 & $4.0\times10^3$ \\
\hline \hline
\end{tabular}
\label{tab:paramII}
\end{center}
\end{table*}

In this scenario, the spectral bumps around 10 TeV are ascribed to the 
cutoff of population I, with a characteristic cutoff rigidity of $\sim60$ TV. 
The spectra become harder again for rigidities higher than $\sim100$ TV,
due to the contribution from population II. The cutoff rigidity of 
population II is about 4 PV, which corresponds to the knee of the
all-particle spectrum. We note that the expected spectrum of p+He of
this model should also show bump-like feature as that seen in the spectra
of protons and Helium. The data from CREAM do show hints of this kind of
feature \cite{2017ApJ...839....5Y}. The preliminary result about the p+He
spectrum by HAWC also shows the bump feature at $\sim30$ TeV
\cite{HAWC-ICRC2019-176}, consistent with the model fittings in this work.
However, the ARGO-WFCTA data show that the knee of the p+He spectrum is 
around 700 TeV, which is lower than the $4\sim8$ PeV obtained in our fittings. 
This is because we use the KASCADE measurements to determine the cutoff 
energy of population II. As shown in Ref.~\cite{2018ChPhC..42g5103G}, 
the fitting to KASCADE data does favor a higher cutoff energy than the 
fitting to ARGO data. Improved measurements of the p+He spectra above 
100 TeV energies are necessary to understand this slight tension.

\subsection{Nearby source(s)}

\begin{figure*}[!htb]
\includegraphics[width=0.45\textwidth]{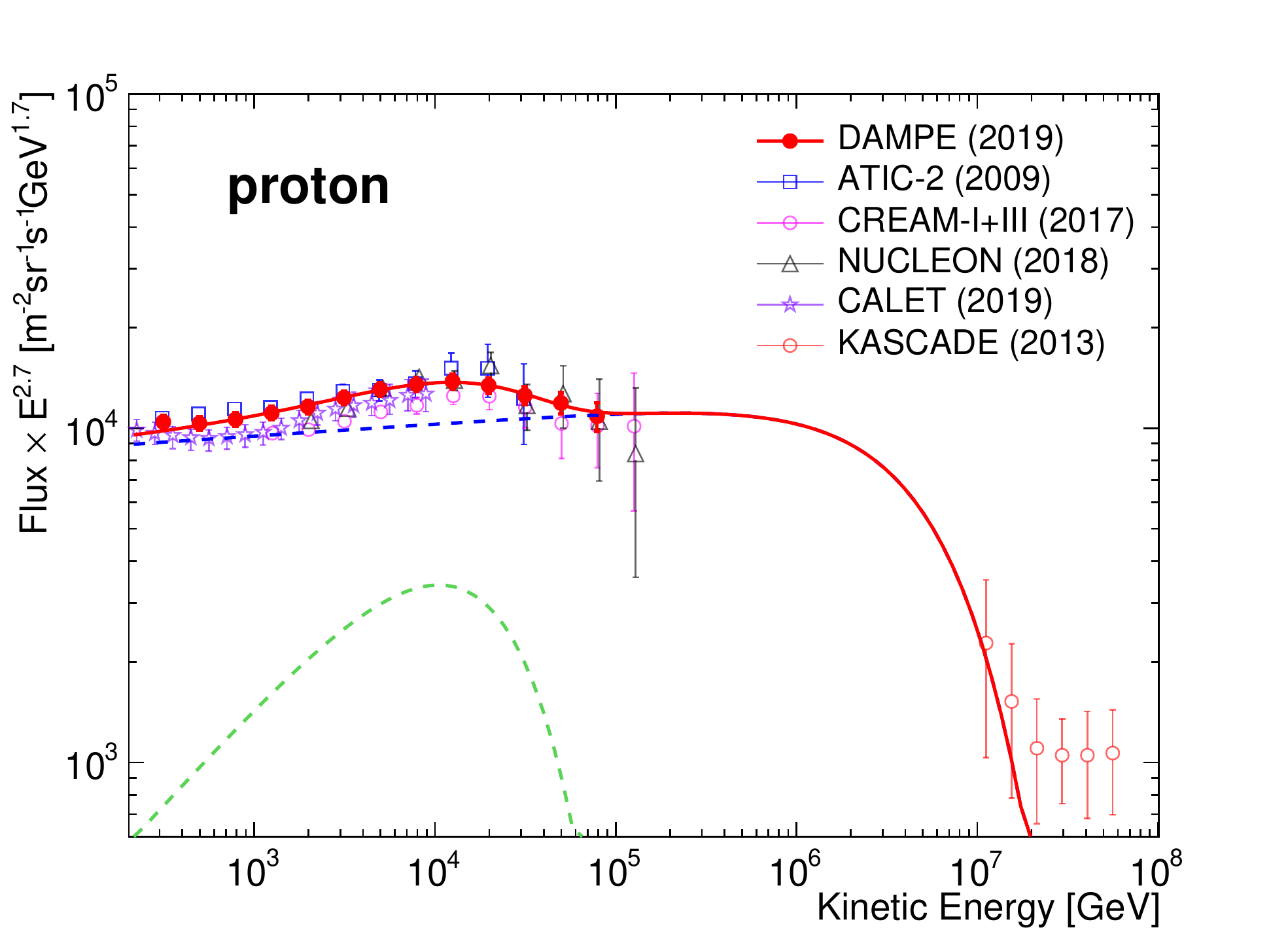}
\includegraphics[width=0.45\textwidth]{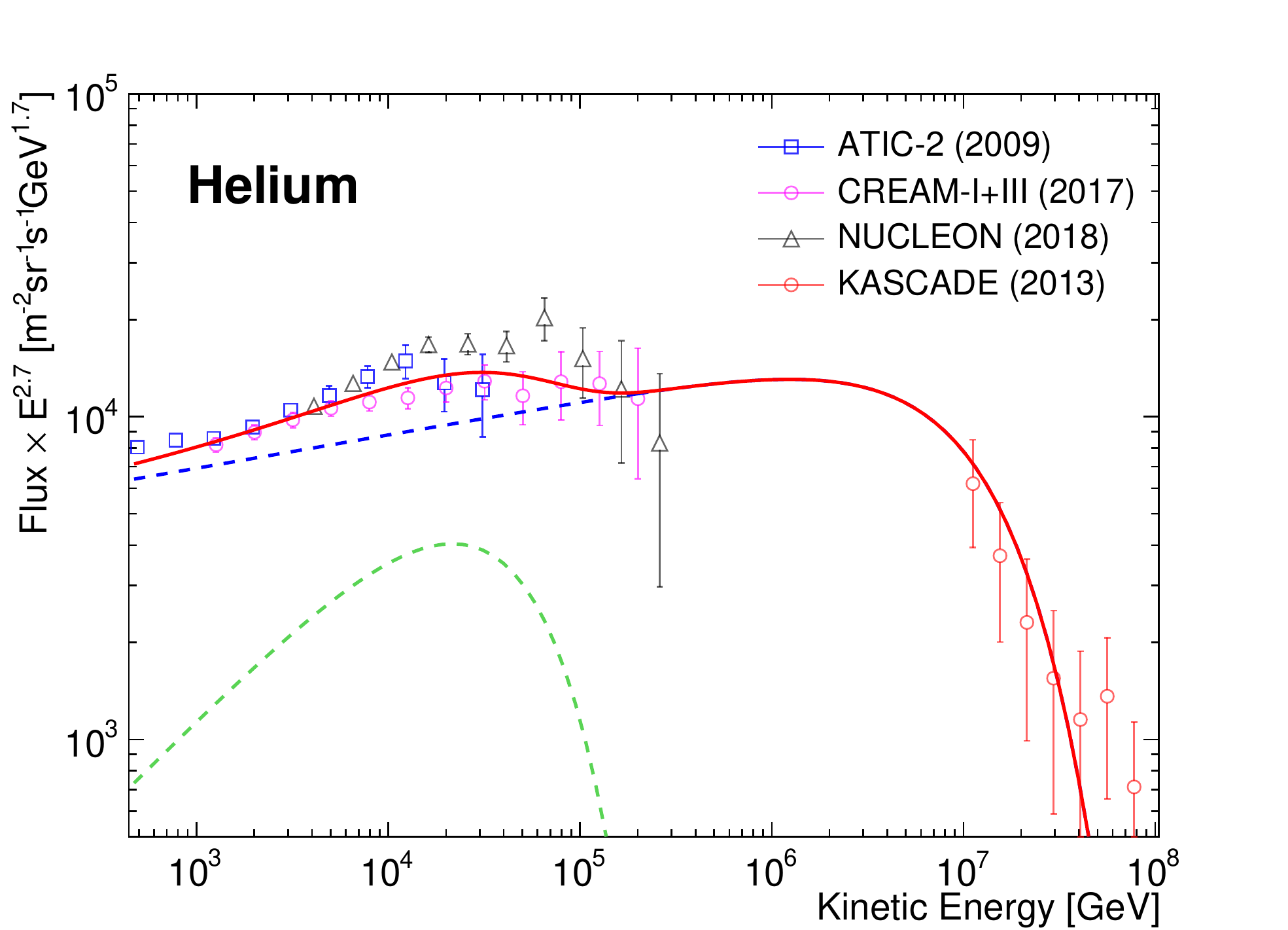}
\includegraphics[width=0.45\textwidth]{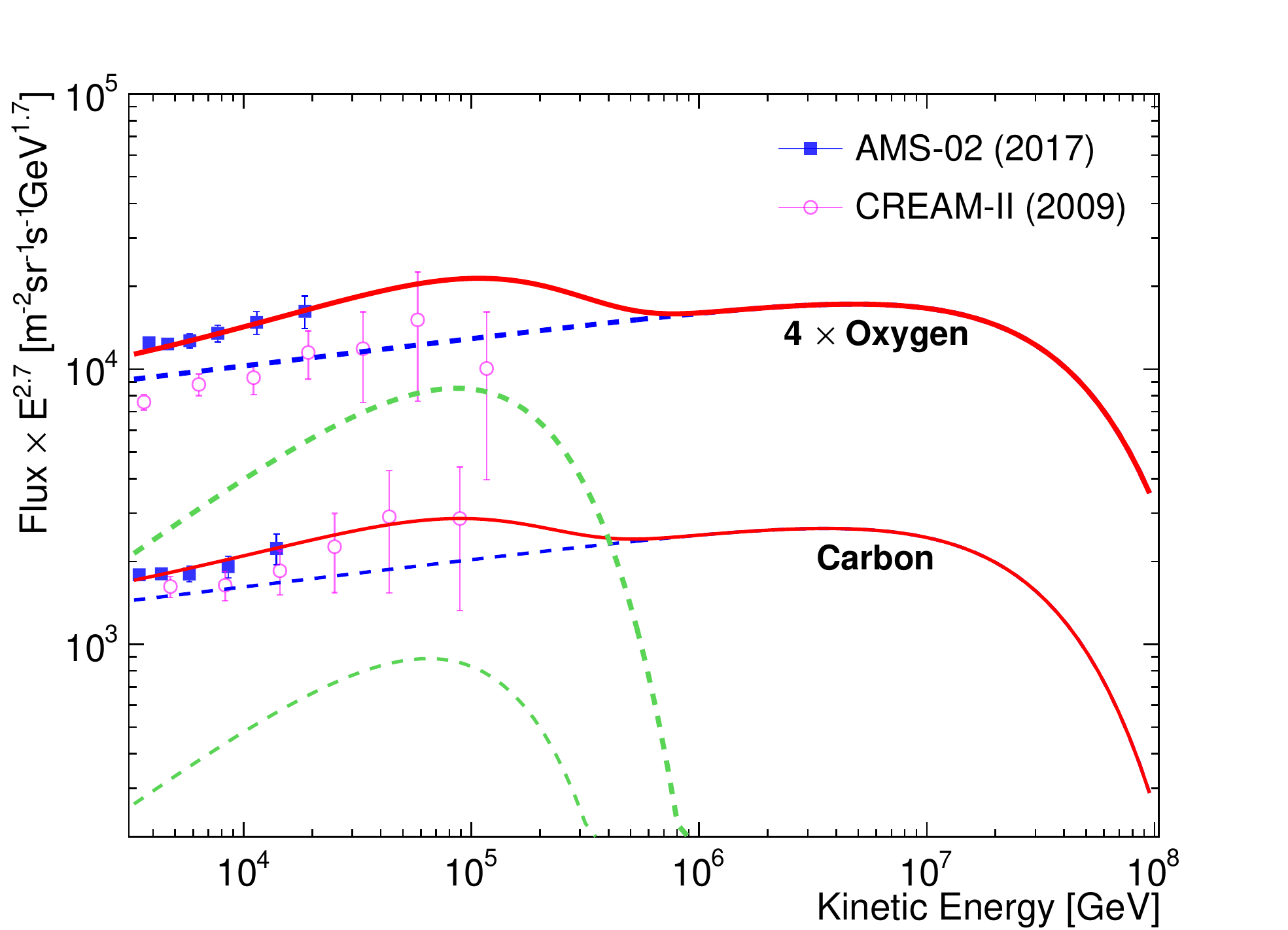}
\includegraphics[width=0.45\textwidth]{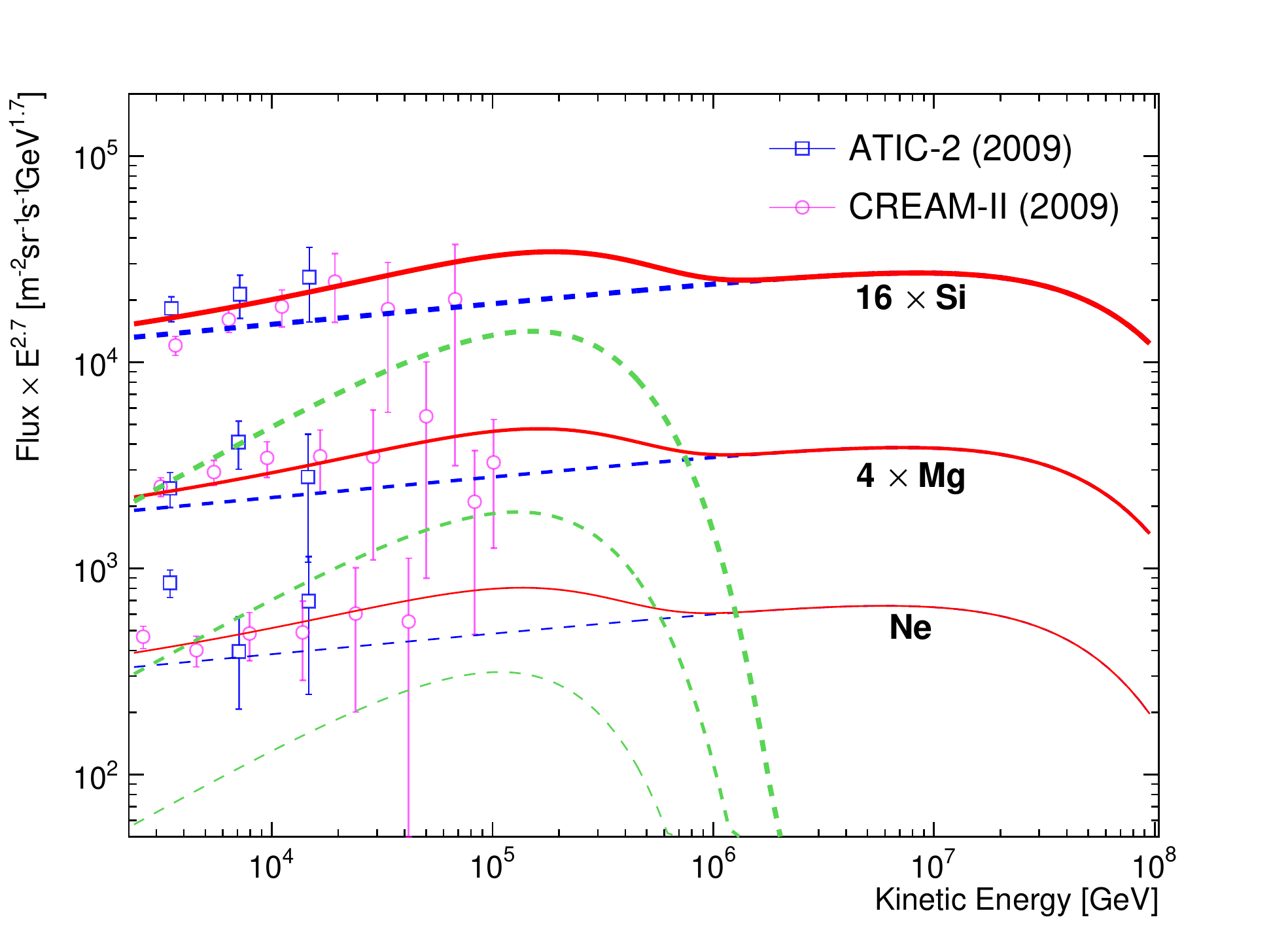}
\includegraphics[width=0.45\textwidth]{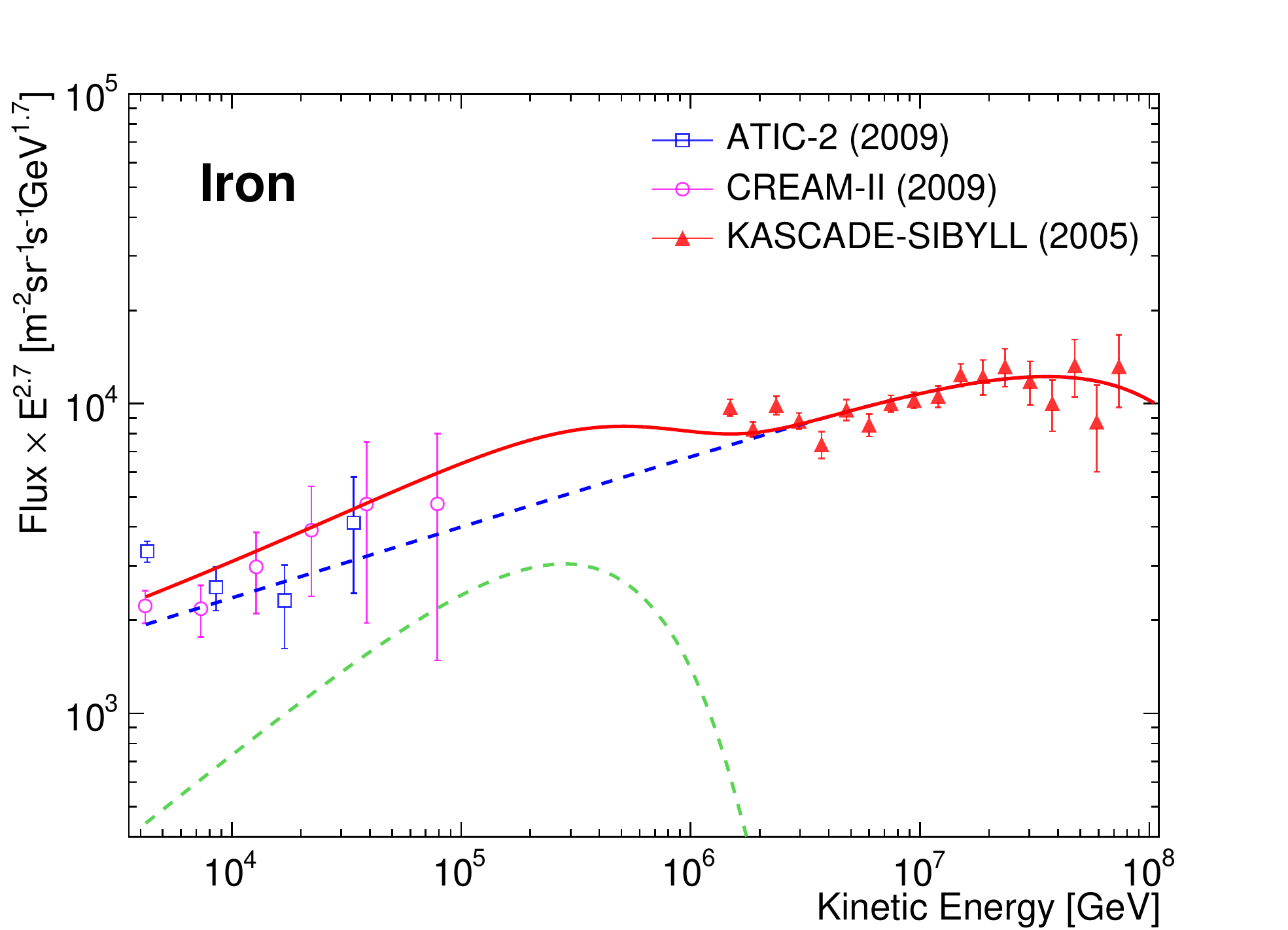}
\includegraphics[width=0.45\textwidth]{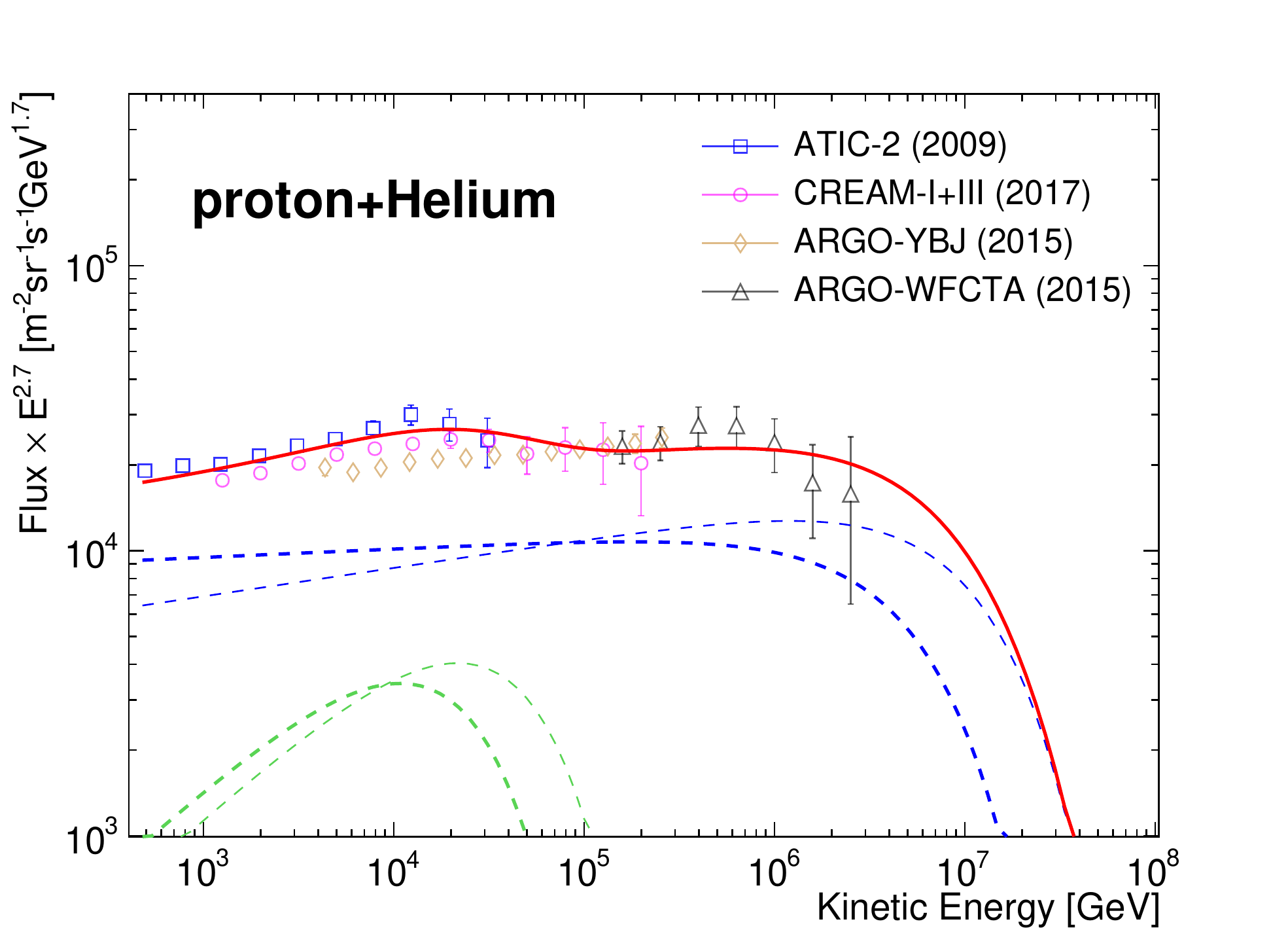}
\includegraphics[width=0.45\textwidth]{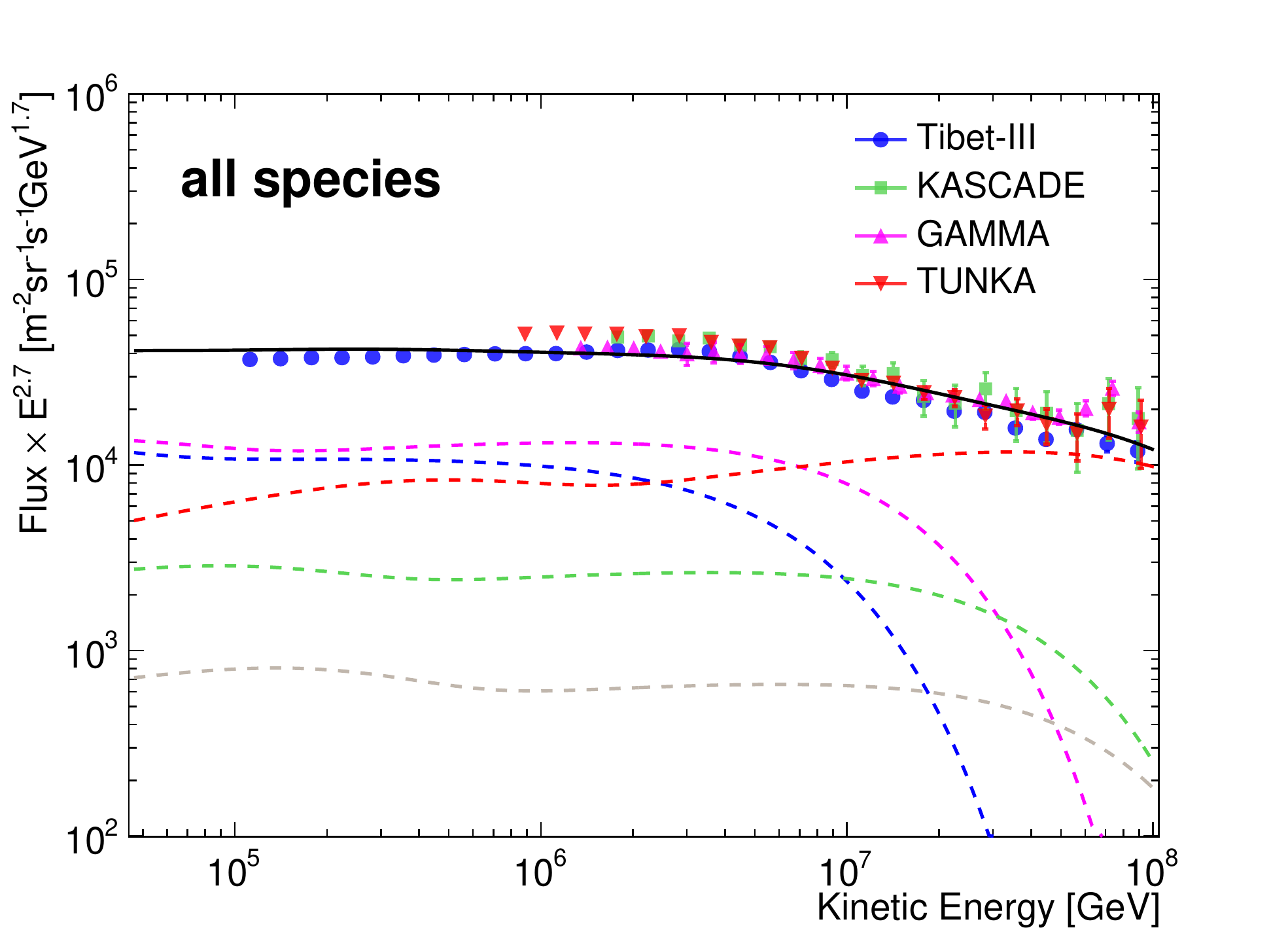}
\caption{Same as Fig.~\ref{fig:specII} but for model B.
\label{fig:specIII}}
\end{figure*}

The other scenario to ascribe these spectral features to the contribution 
of nearby source(s). We assume that the majorities of the observed CR fluxes 
are due to a background component from the population of sources, and a 
nearby source component contributes to the $\sim 10$ TV spectral bumps.
The energy spectra of both the background and the nearby components are
assumed to be exponentially cutoff power-law functions. The fitting 
results are shown in Fig.~\ref{fig:specIII}, with best-fit parameters
compiled in Table \ref{tab:paramIII}. For the nearby source, the spectral
index is about $2.1$ and the cutoff rigidity is about $20$ TV. Note that
in Ref.~\cite{2019JCAP...10..010L} a slightly higher cutoff rigidity of
$\sim 70$ TV was derived to fit the CREAM data. This difference is
probably due to that the DAMPE data is used in the fit here, 
and we neglect the CR propagation in this work. This nearby 
source model (model B) gives comparable goodness-of-fit to the current 
data, compared with model A described in Sec. III A. These two models 
may have slight differences in predicting the spectra between 100 TeV 
and 10 PeV where measurements are lacking. However, we should note that 
such differences may become smaller through adjusting the model parameters.

\begin{table*}[!htb]
\begin{center}
\caption{Spectral parameters of model B.}
\begin{tabular}{cccc|ccc}\hline \hline
        & \multicolumn{3}{c|}{Background} & \multicolumn{3}{c}{Nearby source} \\
Species & $\Phi_{0,i}$ & $\gamma_i$ & $\epsilon_c$ & $\Phi_{0,i}$ & $\gamma_i$ & $\epsilon_c$ \\ 
        & (m$^{-2}$s$^{-1}$sr$^{-1}$TeV$^{-1}$) & & (TeV) & (m$^{-2}$s$^{-1}$sr$^{-1}$TeV$^{-1}$) & & (TeV) \\ \hline
p       & $7.41\times10^{-2}$ & 2.66 & $6.0\times10^3$ & $1.18\times10^{-2}$ & 2.10 & $18$ \\
He      & $5.55\times10^{-2}$ & 2.60 & $6.0\times10^3$ & $9.30\times10^{-3}$ & 2.10 & $18$ \\
C       & $1.02\times10^{-2}$ & 2.60 & $6.0\times10^3$ & $1.10\times10^{-3}$ & 2.10 & $18$ \\
O       & $1.63\times10^{-2}$ & 2.60 & $6.0\times10^3$ & $2.20\times10^{-3}$ & 2.10 & $18$ \\
Ne      & $2.40\times10^{-3}$ & 2.60 & $6.0\times10^3$ & $2.64\times10^{-4}$ & 2.10 & $18$ \\
Mg      & $3.52\times10^{-3}$ & 2.60 & $6.0\times10^3$ & $4.03\times10^{-4}$ & 2.10 & $18$ \\
Si      & $6.08\times10^{-3}$ & 2.60 & $6.0\times10^3$ & $6.37\times10^{-4}$ & 2.10 & $18$ \\
Fe      & $1.16\times10^{-2}$ & 2.48 & $6.0\times10^3$ & $1.28\times10^{-3}$ & 2.10 & $18$ \\
\hline \hline
\end{tabular}
\label{tab:paramIII}
\end{center}
\end{table*}

Nevertheless, there is a potentially significant difference between models
A and B, i.e., the predicted anisotropy pattern of arrival directions of CRs. 
For model A, the predicted large-scale anisotropies of CRs are the same as 
the conventional CR diffusion model with a single component of source 
distribution. The amplitudes of the dipole anisotropies are proportional 
to $E^{\delta}$, where $\delta$ is the energy-dependent slope of the 
diffusion coefficient. The direction of the anisotropy pattern points from 
the Galactic center to the anti-center. These model predictions are, 
however, inconsistent with the measurements of the anisotropies 
\cite{1996ApJ...470..501A,2006Sci...314..439A,2009ApJ...692L.130A,
2016ApJ...826..220A,2017ApJ...836..153A}. Model B can explain the anisotropy 
data well \cite{2015ApJ...809L..23S,2019JCAP...10..010L}. As suggested in 
Ref.~\cite{2019JCAP...10..010L}, a local source located in the direction
that close to Geminga, together with the background source component,
can simultaneously explain the spectral features of CR protons and 
Helium nuclei and the amplitudes and phases of the dipole anisotropies.
Specifically, the nearby source dominates the low-energy ($E<100$ TeV) 
anisotropies with phases being determined by the direction of the source, 
and the background dominates the high-energy ($E>100$ TeV) anisotropies 
with phases pointing from the Galactic center to the anti-center.

\section{Conclusion}

Direct measurements of the CR spectra up to 100 TeV by CREAM, NUCLEON, 
and particularly by DAMPE with high-precision, reveal spectral softenings 
around $\sim10$ TV rigidities. In this work we discuss possible origins 
of these results, taking into account the wide-band measurements of the 
CR energy spectra of various mass groups. We show that employing two 
populations of CR sources with cutoff rigidities of $\sim 60$ TV and 
$\sim 4$ PV can properly fit the measured energy spectra of the main 
species as well as the all-particle spectrum. Alternatively, including 
a nearby source on top of the background component gives similar fitting 
to the spectra. The nearby source model can additionally explain the 
amplitudes and phases of the large-scale anisotropies of CRs, as long as 
the source is located at a proper direction in the sky. It has been found 
that the Geminga supernova remnant may be a promising candidate of such a
local source \cite{2019JCAP...10..010L}.

The revealing of new spectral features of CRs is shown to be able to 
give very interesting implications on the physics of CRs. The measurement
uncertainties of the energy spectra of different mass groups are relatively 
large for energies higher than 100 TeV, due to the low statistics (for
space detection) or the poor composition resolution (for ground-based
detection). The under construction Large High Altitude Air Shower Observatory 
(LHAASO; \cite{2019arXiv190502773B}) and the proposed High Energy
cosmic-Radiation Detection (HERD; \cite{2014SPIE.9144E..0XZ}) facility
onboard the Chinese Space Station are expected to significantly improve 
the precision of CR spectral measurements. Particularly, the measurements 
of anisotropies of different mass groups by LHAASO will be essentially 
helpful in understanding the spectral softening features, the knee 
structures, and the origin of CRs in general. 

\acknowledgments
This work is supported by the National Key Research and Development Program 
of China (No. 2016YFA0400200), the National Natural Science Foundation of 
China (Nos. 11722328, 11525313, U1738205, 11851305), and the 100 Talents 
Program of Chinese Academy of Sciences.


\begin{thebibliography}{68}
\expandafter\ifx\csname natexlab\endcsname\relax\def\natexlab#1{#1}\fi
\expandafter\ifx\csname bibnamefont\endcsname\relax
  \def\bibnamefont#1{#1}\fi
\expandafter\ifx\csname bibfnamefont\endcsname\relax
  \def\bibfnamefont#1{#1}\fi
\expandafter\ifx\csname citenamefont\endcsname\relax
  \def\citenamefont#1{#1}\fi
\expandafter\ifx\csname url\endcsname\relax
  \def\url#1{\texttt{#1}}\fi
\expandafter\ifx\csname urlprefix\endcsname\relax\def\urlprefix{URL }\fi
\providecommand{\bibinfo}[2]{#2}
\providecommand{\eprint}[2][]{\url{#2}}

\bibitem[{\citenamefont{{Aguilar} et~al.}(2016)\citenamefont{{Aguilar}, {Aisa},
  {Alvino}, {Ambrosi}, {Andeen}, {Arruda}, {Attig}, {Azzarello}, {Bachlechner},
  {Barao} et~al.}}]{2016PhRvL.117w1102A}
\bibinfo{author}{\bibfnamefont{M.}~\bibnamefont{{Aguilar}}},
  \bibnamefont{et~al.}, \bibinfo{journal}{\prl} \textbf{\bibinfo{volume}{117}},
  \bibinfo{eid}{231102} (\bibinfo{year}{2016}).

\bibitem[{\citenamefont{{Panov} et~al.}(2009)\citenamefont{{Panov}, {Adams},
  {Ahn}, {Bashinzhagyan}, {Watts}, {Wefel}, {Wu}, {Ganel}, {Guzik}, {Zatsepin}
  et~al.}}]{2009BRASP..73..564P}
\bibinfo{author}{\bibfnamefont{A.~D.} \bibnamefont{{Panov}}},
  \bibnamefont{et~al.}, \bibinfo{journal}{Bulletin of the Russian Academy of
  Science, Phys.} \textbf{\bibinfo{volume}{73}}, \bibinfo{pages}{564}
  (\bibinfo{year}{2009}), \eprint{1101.3246}.

\bibitem[{\citenamefont{{Ahn} et~al.}(2010)\citenamefont{{Ahn}, {Allison},
  {Bagliesi}, {Beatty}, {Bigongiari}, {Childers}, {Conklin}, {Coutu},
  {DuVernois}, {Ganel} et~al.}}]{2010ApJ...714L..89A}
\bibinfo{author}{\bibfnamefont{H.~S.} \bibnamefont{{Ahn}}},
  \bibnamefont{et~al.}, \bibinfo{journal}{\apjl}
  \textbf{\bibinfo{volume}{714}}, \bibinfo{pages}{L89} (\bibinfo{year}{2010}),
  \eprint{1004.1123}.

\bibitem[{\citenamefont{{Adriani} et~al.}(2011)\citenamefont{{Adriani},
  {Barbarino}, {Bazilevskaya}, {Bellotti}, {Boezio}, {Bogomolov}, {Bonechi},
  {Bongi}, {Bonvicini}, {Borisov} et~al.}}]{2011Sci...332...69A}
\bibinfo{author}{\bibfnamefont{O.}~\bibnamefont{{Adriani}}},
  \bibnamefont{et~al.}, \bibinfo{journal}{Science}
  \textbf{\bibinfo{volume}{332}}, \bibinfo{pages}{69} (\bibinfo{year}{2011}),
  \eprint{1103.4055}.

\bibitem[{\citenamefont{{Aguilar}
  et~al.}(2015{\natexlab{a}})\citenamefont{{Aguilar}, {Aisa}, {Alpat},
  {Alvino}, {Ambrosi}, {Andeen}, {Arruda}, {Attig}, {Azzarello}, {Bachlechner}
  et~al.}}]{2015PhRvL.114q1103A}
\bibinfo{author}{\bibfnamefont{M.}~\bibnamefont{{Aguilar}}},
  \bibnamefont{et~al.}, \bibinfo{journal}{\prl} \textbf{\bibinfo{volume}{114}},
  \bibinfo{eid}{171103} (\bibinfo{year}{2015}{\natexlab{a}}).

\bibitem[{\citenamefont{{Aguilar}
  et~al.}(2015{\natexlab{b}})\citenamefont{{Aguilar}, {Aisa}, {Alpat},
  {Alvino}, {Ambrosi}, {Andeen}, {Arruda}, {Attig}, {Azzarello}, {Bachlechner}
  et~al.}}]{2015PhRvL.115u1101A}
\bibinfo{author}{\bibfnamefont{M.}~\bibnamefont{{Aguilar}}},
  \bibnamefont{et~al.}, \bibinfo{journal}{\prl} \textbf{\bibinfo{volume}{115}},
  \bibinfo{eid}{211101} (\bibinfo{year}{2015}{\natexlab{b}}).

\bibitem[{\citenamefont{{Aguilar} et~al.}(2017)\citenamefont{{Aguilar}, {Ali
  Cavasonza}, {Alpat}, {Ambrosi}, {Arruda}, {Attig}, {Aupetit}, {Azzarello},
  {Bachlechner}, {Barao} et~al.}}]{2017PhRvL.119y1101A}
\bibinfo{author}{\bibfnamefont{M.}~\bibnamefont{{Aguilar}}},
  \bibnamefont{et~al.}, \bibinfo{journal}{\prl} \textbf{\bibinfo{volume}{119}},
  \bibinfo{eid}{251101} (\bibinfo{year}{2017}).

\bibitem[{\citenamefont{{Adriani} et~al.}(2019)\citenamefont{{Adriani},
  {Akaike}, {Asano}, {Asaoka}, {Bagliesi}, {Berti}, {Bigongiari}, {Binns},
  {Bonechi}, {Bongi} et~al.}}]{2019PhRvL.122r1102A}
\bibinfo{author}{\bibfnamefont{O.}~\bibnamefont{{Adriani}}},
  \bibnamefont{et~al.}, \bibinfo{journal}{\prl} \textbf{\bibinfo{volume}{122}},
  \bibinfo{eid}{181102} (\bibinfo{year}{2019}), \eprint{1905.04229}.

\bibitem[{\citenamefont{{Ohira} and {Ioka}}(2011)}]{2011ApJ...729L..13O}
\bibinfo{author}{\bibfnamefont{Y.}~\bibnamefont{{Ohira}}} \bibnamefont{and}
  \bibinfo{author}{\bibfnamefont{K.}~\bibnamefont{{Ioka}}},
  \bibinfo{journal}{\apjl} \textbf{\bibinfo{volume}{729}}, \bibinfo{pages}{L13}
  (\bibinfo{year}{2011}), \eprint{1011.4405}.

\bibitem[{\citenamefont{{Yuan} et~al.}(2011)\citenamefont{{Yuan}, {Zhang}, and
  {Bi}}}]{2011PhRvD..84d3002Y}
\bibinfo{author}{\bibfnamefont{Q.}~\bibnamefont{{Yuan}}},
  \bibinfo{author}{\bibfnamefont{B.}~\bibnamefont{{Zhang}}}, \bibnamefont{and}
  \bibinfo{author}{\bibfnamefont{X.-J.} \bibnamefont{{Bi}}},
  \bibinfo{journal}{\prd} \textbf{\bibinfo{volume}{84}},
  \bibinfo{pages}{043002} (\bibinfo{year}{2011}), \eprint{1104.3357}.

\bibitem[{\citenamefont{{Vladimirov} et~al.}(2012)\citenamefont{{Vladimirov},
  {J{\'o}hannesson}, {Moskalenko}, and {Porter}}}]{2012ApJ...752...68V}
\bibinfo{author}{\bibfnamefont{A.~E.} \bibnamefont{{Vladimirov}}},
  \bibinfo{author}{\bibfnamefont{G.}~\bibnamefont{{J{\'o}hannesson}}},
  \bibinfo{author}{\bibfnamefont{I.~V.} \bibnamefont{{Moskalenko}}},
  \bibnamefont{and} \bibinfo{author}{\bibfnamefont{T.~A.}
  \bibnamefont{{Porter}}}, \bibinfo{journal}{\apj}
  \textbf{\bibinfo{volume}{752}}, \bibinfo{eid}{68} (\bibinfo{year}{2012}),
  \eprint{1108.1023}.

\bibitem[{\citenamefont{{Erlykin} and
  {Wolfendale}}(2012)}]{2012APh....35..449E}
\bibinfo{author}{\bibfnamefont{A.~D.} \bibnamefont{{Erlykin}}}
  \bibnamefont{and} \bibinfo{author}{\bibfnamefont{A.~W.}
  \bibnamefont{{Wolfendale}}}, \bibinfo{journal}{Astroparticle Physics}
  \textbf{\bibinfo{volume}{35}}, \bibinfo{pages}{449} (\bibinfo{year}{2012}).

\bibitem[{\citenamefont{{Thoudam} and
  {H{\"o}randel}}(2012)}]{2012MNRAS.421.1209T}
\bibinfo{author}{\bibfnamefont{S.}~\bibnamefont{{Thoudam}}} \bibnamefont{and}
  \bibinfo{author}{\bibfnamefont{J.~R.} \bibnamefont{{H{\"o}randel}}},
  \bibinfo{journal}{\mnras} \textbf{\bibinfo{volume}{421}},
  \bibinfo{pages}{1209} (\bibinfo{year}{2012}), \eprint{1112.3020}.

\bibitem[{\citenamefont{{Bernard} et~al.}(2013)\citenamefont{{Bernard},
  {Delahaye}, {Keum}, {Liu}, {Salati}, and {Taillet}}}]{2013A&A...555A..48B}
\bibinfo{author}{\bibfnamefont{G.}~\bibnamefont{{Bernard}}},
  \bibinfo{author}{\bibfnamefont{T.}~\bibnamefont{{Delahaye}}},
  \bibinfo{author}{\bibfnamefont{Y.-Y.} \bibnamefont{{Keum}}},
  \bibinfo{author}{\bibfnamefont{W.}~\bibnamefont{{Liu}}},
  \bibinfo{author}{\bibfnamefont{P.}~\bibnamefont{{Salati}}}, \bibnamefont{and}
  \bibinfo{author}{\bibfnamefont{R.}~\bibnamefont{{Taillet}}},
  \bibinfo{journal}{\aap} \textbf{\bibinfo{volume}{555}}, \bibinfo{eid}{A48}
  (\bibinfo{year}{2013}), \eprint{1207.4670}.

\bibitem[{\citenamefont{{Liu} et~al.}(2017)\citenamefont{{Liu}, {Bi}, {Lin},
  {Wang}, and {Yin}}}]{2017PhRvD..96b3006L}
\bibinfo{author}{\bibfnamefont{W.}~\bibnamefont{{Liu}}},
  \bibinfo{author}{\bibfnamefont{X.-J.} \bibnamefont{{Bi}}},
  \bibinfo{author}{\bibfnamefont{S.-J.} \bibnamefont{{Lin}}},
  \bibinfo{author}{\bibfnamefont{B.-B.} \bibnamefont{{Wang}}},
  \bibnamefont{and} \bibinfo{author}{\bibfnamefont{P.-F.} \bibnamefont{{Yin}}},
  \bibinfo{journal}{\prd} \textbf{\bibinfo{volume}{96}}, \bibinfo{eid}{023006}
  (\bibinfo{year}{2017}), \eprint{1611.09118}.

\bibitem[{\citenamefont{{Ptuskin} et~al.}(2013)\citenamefont{{Ptuskin},
  {Zirakashvili}, and {Seo}}}]{2013ApJ...763...47P}
\bibinfo{author}{\bibfnamefont{V.}~\bibnamefont{{Ptuskin}}},
  \bibinfo{author}{\bibfnamefont{V.}~\bibnamefont{{Zirakashvili}}},
  \bibnamefont{and} \bibinfo{author}{\bibfnamefont{E.-S.} \bibnamefont{{Seo}}},
  \bibinfo{journal}{\apj} \textbf{\bibinfo{volume}{763}}, \bibinfo{eid}{47}
  (\bibinfo{year}{2013}), \eprint{1212.0381}.

\bibitem[{\citenamefont{{Thoudam} and
  {H{\"o}randel}}(2014)}]{2014A&A...567A..33T}
\bibinfo{author}{\bibfnamefont{S.}~\bibnamefont{{Thoudam}}} \bibnamefont{and}
  \bibinfo{author}{\bibfnamefont{J.~R.} \bibnamefont{{H{\"o}randel}}},
  \bibinfo{journal}{\aap} \textbf{\bibinfo{volume}{567}}, \bibinfo{eid}{A33}
  (\bibinfo{year}{2014}), \eprint{1404.3630}.

\bibitem[{\citenamefont{{Zhang} et~al.}(2017)\citenamefont{{Zhang}, {Liu}, and
  {Yuan}}}]{2017ApJ...844L...3Z}
\bibinfo{author}{\bibfnamefont{Y.}~\bibnamefont{{Zhang}}},
  \bibinfo{author}{\bibfnamefont{S.}~\bibnamefont{{Liu}}}, \bibnamefont{and}
  \bibinfo{author}{\bibfnamefont{Q.}~\bibnamefont{{Yuan}}},
  \bibinfo{journal}{\apjl} \textbf{\bibinfo{volume}{844}}, \bibinfo{eid}{L3}
  (\bibinfo{year}{2017}), \eprint{1707.00262}.

\bibitem[{\citenamefont{{Tomassetti}}(2012)}]{2012ApJ...752L..13T}
\bibinfo{author}{\bibfnamefont{N.}~\bibnamefont{{Tomassetti}}},
  \bibinfo{journal}{\apjl} \textbf{\bibinfo{volume}{752}}, \bibinfo{eid}{L13}
  (\bibinfo{year}{2012}), \eprint{1204.4492}.

\bibitem[{\citenamefont{{Blasi} et~al.}(2012)\citenamefont{{Blasi}, {Amato},
  and {Serpico}}}]{2012PhRvL.109f1101B}
\bibinfo{author}{\bibfnamefont{P.}~\bibnamefont{{Blasi}}},
  \bibinfo{author}{\bibfnamefont{E.}~\bibnamefont{{Amato}}}, \bibnamefont{and}
  \bibinfo{author}{\bibfnamefont{P.~D.} \bibnamefont{{Serpico}}},
  \bibinfo{journal}{\prl} \textbf{\bibinfo{volume}{109}}, \bibinfo{eid}{061101}
  (\bibinfo{year}{2012}), \eprint{1207.3706}.

\bibitem[{\citenamefont{{Tomassetti} and {Donato}}(2015)}]{2015ApJ...803L..15T}
\bibinfo{author}{\bibfnamefont{N.}~\bibnamefont{{Tomassetti}}}
  \bibnamefont{and} \bibinfo{author}{\bibfnamefont{F.}~\bibnamefont{{Donato}}},
  \bibinfo{journal}{\apjl} \textbf{\bibinfo{volume}{803}}, \bibinfo{eid}{L15}
  (\bibinfo{year}{2015}), \eprint{1502.06150}.

\bibitem[{\citenamefont{{Taylor} and {Giacinti}}(2017)}]{2017PhRvD..95b3001T}
\bibinfo{author}{\bibfnamefont{A.~M.} \bibnamefont{{Taylor}}} \bibnamefont{and}
  \bibinfo{author}{\bibfnamefont{G.}~\bibnamefont{{Giacinti}}},
  \bibinfo{journal}{\prd} \textbf{\bibinfo{volume}{95}}, \bibinfo{eid}{023001}
  (\bibinfo{year}{2017}), \eprint{1607.08862}.

\bibitem[{\citenamefont{{Jin} et~al.}(2016)\citenamefont{{Jin}, {Guo}, and
  {Hu}}}]{2016ChPhC..40a5101J}
\bibinfo{author}{\bibfnamefont{C.}~\bibnamefont{{Jin}}},
  \bibinfo{author}{\bibfnamefont{Y.-Q.} \bibnamefont{{Guo}}}, \bibnamefont{and}
  \bibinfo{author}{\bibfnamefont{H.-B.} \bibnamefont{{Hu}}},
  \bibinfo{journal}{Chinese Physics C} \textbf{\bibinfo{volume}{40}},
  \bibinfo{eid}{015101} (\bibinfo{year}{2016}), \eprint{1504.06903}.

\bibitem[{\citenamefont{{Guo} et~al.}(2016)\citenamefont{{Guo}, {Tian}, and
  {Jin}}}]{2016ApJ...819...54G}
\bibinfo{author}{\bibfnamefont{Y.-Q.} \bibnamefont{{Guo}}},
  \bibinfo{author}{\bibfnamefont{Z.}~\bibnamefont{{Tian}}}, \bibnamefont{and}
  \bibinfo{author}{\bibfnamefont{C.}~\bibnamefont{{Jin}}},
  \bibinfo{journal}{\apj} \textbf{\bibinfo{volume}{819}}, \bibinfo{eid}{54}
  (\bibinfo{year}{2016}).

\bibitem[{\citenamefont{{Guo} and
  {Yuan}}(2018{\natexlab{a}})}]{2018PhRvD..97f3008G}
\bibinfo{author}{\bibfnamefont{Y.-Q.} \bibnamefont{{Guo}}} \bibnamefont{and}
  \bibinfo{author}{\bibfnamefont{Q.}~\bibnamefont{{Yuan}}},
  \bibinfo{journal}{\prd} \textbf{\bibinfo{volume}{97}}, \bibinfo{eid}{063008}
  (\bibinfo{year}{2018}{\natexlab{a}}), \eprint{1801.05904}.

\bibitem[{\citenamefont{{Liu} et~al.}(2018)\citenamefont{{Liu}, {Yao}, and
  {Guo}}}]{2018ApJ...869..176L}
\bibinfo{author}{\bibfnamefont{W.}~\bibnamefont{{Liu}}},
  \bibinfo{author}{\bibfnamefont{Y.-h.} \bibnamefont{{Yao}}}, \bibnamefont{and}
  \bibinfo{author}{\bibfnamefont{Y.-Q.} \bibnamefont{{Guo}}},
  \bibinfo{journal}{\apj} \textbf{\bibinfo{volume}{869}}, \bibinfo{eid}{176}
  (\bibinfo{year}{2018}), \eprint{1802.03602}.

\bibitem[{\citenamefont{{Aguilar} et~al.}(2018)\citenamefont{{Aguilar}, {Ali
  Cavasonza}, {Alpat}, {Ambrosi}, {Arruda}, {Attig}, {Aupetit}, {Azzarello},
  {Bachlechner}, {Barao} et~al.}}]{2018PhRvL.120b1101A}
\bibinfo{author}{\bibfnamefont{M.}~\bibnamefont{{Aguilar}}},
  \bibnamefont{et~al.}, \bibinfo{journal}{\prl} \textbf{\bibinfo{volume}{120}},
  \bibinfo{eid}{021101} (\bibinfo{year}{2018}).

\bibitem[{\citenamefont{{G{\'e}nolini}
  et~al.}(2017)\citenamefont{{G{\'e}nolini}, {Serpico}, {Boudaud}, {Caroff},
  {Poulin}, {Derome}, {Lavalle}, {Maurin}, {Poireau}, {Rosier}
  et~al.}}]{2017PhRvL.119x1101G}
\bibinfo{author}{\bibfnamefont{Y.}~\bibnamefont{{G{\'e}nolini}}},
  \bibnamefont{et~al.}, \bibinfo{journal}{\prl} \textbf{\bibinfo{volume}{119}},
  \bibinfo{eid}{241101} (\bibinfo{year}{2017}), \eprint{1706.09812}.

\bibitem[{\citenamefont{{Yuan} et~al.}(2018)\citenamefont{{Yuan}, {Zhu}, {Bi},
  and {Wei}}}]{2018arXiv181003141Y}
\bibinfo{author}{\bibfnamefont{Q.}~\bibnamefont{{Yuan}}},
  \bibinfo{author}{\bibfnamefont{C.-R.} \bibnamefont{{Zhu}}},
  \bibinfo{author}{\bibfnamefont{X.-J.} \bibnamefont{{Bi}}}, \bibnamefont{and}
  \bibinfo{author}{\bibfnamefont{D.-M.} \bibnamefont{{Wei}}},
  \bibinfo{journal}{arXiv e-prints}  (\bibinfo{year}{2018}),
  \eprint{1810.03141}.

\bibitem[{\citenamefont{{Niu} et~al.}(2018)\citenamefont{{Niu}, {Li}, and
  {Xue}}}]{2018arXiv181009301N}
\bibinfo{author}{\bibfnamefont{J.-S.} \bibnamefont{{Niu}}},
  \bibinfo{author}{\bibfnamefont{T.}~\bibnamefont{{Li}}}, \bibnamefont{and}
  \bibinfo{author}{\bibfnamefont{H.-F.} \bibnamefont{{Xue}}},
  \bibinfo{journal}{arXiv e-prints}  (\bibinfo{year}{2018}),
  \eprint{1810.09301}.

\bibitem[{\citenamefont{{Yoon} et~al.}(2017)\citenamefont{{Yoon}, {Anderson},
  {Barrau}, {Conklin}, {Coutu}, {Derome}, {Han}, {Jeon}, {Kim}, {Kim}
  et~al.}}]{2017ApJ...839....5Y}
\bibinfo{author}{\bibfnamefont{Y.~S.} \bibnamefont{{Yoon}}},
  \bibnamefont{et~al.}, \bibinfo{journal}{\apj} \textbf{\bibinfo{volume}{839}},
  \bibinfo{eid}{5} (\bibinfo{year}{2017}), \eprint{1704.02512}.

\bibitem[{\citenamefont{{Atkin} et~al.}(2018)\citenamefont{{Atkin}, {Bulatov},
  {Dorokhov}, {Gorbunov}, {Filippov}, {Grebenyuk}, {Karmanov}, {Kovalev},
  {Kudryashov}, {Kurganov} et~al.}}]{2018JETPL.108....5A}
\bibinfo{author}{\bibfnamefont{E.}~\bibnamefont{{Atkin}}},
  \bibnamefont{et~al.}, \bibinfo{journal}{Soviet Journal of Experimental and
  Theoretical Physics Letters} \textbf{\bibinfo{volume}{108}},
  \bibinfo{pages}{5} (\bibinfo{year}{2018}), \eprint{1805.07119}.

\bibitem[{\citenamefont{{Chang}}(2014)}]{ChangJin:550}
\bibinfo{author}{\bibfnamefont{J.}~\bibnamefont{{Chang}}},
  \bibinfo{journal}{Chinese Journal of Space Science}
  \textbf{\bibinfo{volume}{34}}, \bibinfo{pages}{550} (\bibinfo{year}{2014}).

\bibitem[{\citenamefont{{Chang} et~al.}(2017)\citenamefont{{Chang}, {Ambrosi},
  {An}, {Asfandiyarov}, {Azzarello}, {Bernardini}, {Bertucci}, {Cai},
  {Caragiulo}, {Chen} et~al.}}]{2017APh....95....6C}
\bibinfo{author}{\bibfnamefont{J.}~\bibnamefont{{Chang}}},
  \bibnamefont{et~al.}, \bibinfo{journal}{Astroparticle Physics}
  \textbf{\bibinfo{volume}{95}}, \bibinfo{pages}{6} (\bibinfo{year}{2017}),
  \eprint{1706.08453}.

\bibitem[{\citenamefont{{An} et~al.}(2019)\citenamefont{{An}, {Asfandiyarov},
  {Azzarello}, {Bernardini}, {Bi}, {Cai}, {Chang}, {Chen}, {Chen}, {Chen}
  et~al.}}]{2019arXiv190912860A}
\bibinfo{author}{\bibfnamefont{Q.}~\bibnamefont{{An}}}, \bibnamefont{et~al.},
  \bibinfo{journal}{Science Advances} \textbf{\bibinfo{volume}{5}},
  \bibinfo{eid}{eaax3793} (\bibinfo{year}{2019}), \eprint{1909.12860}.

\bibitem[{\citenamefont{{Antoni} et~al.}(2005)\citenamefont{{Antoni}, {Apel},
  {Badea}, {Bekk}, {Bercuci}, {Bl{\"u}mer}, {Bozdog}, {Brancus},
  {Chilingarian}, {Daumiller} et~al.}}]{2005APh....24....1A}
\bibinfo{author}{\bibfnamefont{T.}~\bibnamefont{{Antoni}}},
  \bibnamefont{et~al.}, \bibinfo{journal}{Astropart. Phys.}
  \textbf{\bibinfo{volume}{24}}, \bibinfo{pages}{1} (\bibinfo{year}{2005}),
  \eprint{astro-ph/0505413}.

\bibitem[{\citenamefont{{Korosteleva} et~al.}(2007)\citenamefont{{Korosteleva},
  {Prosin}, {Kuzmichev}, and {Navarra}}}]{2007NuPhS.165...74K}
\bibinfo{author}{\bibfnamefont{E.~E.} \bibnamefont{{Korosteleva}}},
  \bibinfo{author}{\bibfnamefont{V.~V.} \bibnamefont{{Prosin}}},
  \bibinfo{author}{\bibfnamefont{L.~A.} \bibnamefont{{Kuzmichev}}},
  \bibnamefont{and}
  \bibinfo{author}{\bibfnamefont{G.}~\bibnamefont{{Navarra}}},
  \bibinfo{journal}{Nuclear Physics B Proceedings Supplements}
  \textbf{\bibinfo{volume}{165}}, \bibinfo{pages}{74} (\bibinfo{year}{2007}).

\bibitem[{\citenamefont{{Amenomori} et~al.}(2008)\citenamefont{{Amenomori},
  {Bi}, {Chen}, {Cui}, {Danzengluobu}, {Ding}, {Ding}, {Fan}, {Feng}, {Feng}
  et~al.}}]{2008ApJ...678.1165A}
\bibinfo{author}{\bibfnamefont{M.}~\bibnamefont{{Amenomori}}},
  \bibnamefont{et~al.}, \bibinfo{journal}{\apj} \textbf{\bibinfo{volume}{678}},
  \bibinfo{pages}{1165} (\bibinfo{year}{2008}), \eprint{0801.1803}.

\bibitem[{\citenamefont{{Garyaka} et~al.}(2008)\citenamefont{{Garyaka},
  {Martirosov}, {Ter-Antonyan}, {Erlykin}, {Nikolskaya}, {Gallant}, {Jones},
  and {Procureur}}}]{2008JPhG...35k5201G}
\bibinfo{author}{\bibfnamefont{A.~P.} \bibnamefont{{Garyaka}}},
  \bibinfo{author}{\bibfnamefont{R.~M.} \bibnamefont{{Martirosov}}},
  \bibinfo{author}{\bibfnamefont{S.~V.} \bibnamefont{{Ter-Antonyan}}},
  \bibinfo{author}{\bibfnamefont{A.~D.} \bibnamefont{{Erlykin}}},
  \bibinfo{author}{\bibfnamefont{N.~M.} \bibnamefont{{Nikolskaya}}},
  \bibinfo{author}{\bibfnamefont{Y.~A.} \bibnamefont{{Gallant}}},
  \bibinfo{author}{\bibfnamefont{L.~W.} \bibnamefont{{Jones}}},
  \bibnamefont{and}
  \bibinfo{author}{\bibfnamefont{J.}~\bibnamefont{{Procureur}}},
  \bibinfo{journal}{Journal of Physics G Nuclear Physics}
  \textbf{\bibinfo{volume}{35}}, \bibinfo{eid}{115201} (\bibinfo{year}{2008}),
  \eprint{0808.1421}.

\bibitem[{\citenamefont{{Tibet As{$\gamma$} Collaboration}
  et~al.}(2006)\citenamefont{{Tibet As{$\gamma$} Collaboration}, {Amenomori},
  {Ayabe}, {Chen}, {Cui}, {Danzengluobu}, {Ding}, {Ding}, {Feng}, {Feng}
  et~al.}}]{2006PhLB..632...58T}
\bibinfo{author}{\bibnamefont{{Tibet As{$\gamma$} Collaboration}}},
  \bibnamefont{et~al.}, \bibinfo{journal}{Physics Letters B}
  \textbf{\bibinfo{volume}{632}}, \bibinfo{pages}{58} (\bibinfo{year}{2006}),
  \eprint{astro-ph/0511469}.

\bibitem[{\citenamefont{{Bartoli}
  et~al.}(2015{\natexlab{a}})\citenamefont{{Bartoli}, {Bernardini}, {Bi},
  {Cao}, {Catalanotti}, {Chen}, {Chen}, {Cui}, {Dai}, {D'Amone}
  et~al.}}]{2015PhRvD..92i2005B}
\bibinfo{author}{\bibfnamefont{B.}~\bibnamefont{{Bartoli}}},
  \bibnamefont{et~al.}, \bibinfo{journal}{\prd} \textbf{\bibinfo{volume}{92}},
  \bibinfo{eid}{092005} (\bibinfo{year}{2015}{\natexlab{a}}),
  \eprint{1502.03164}.

\bibitem[{\citenamefont{{Arteaga-Velazquez} and
  {Alvarez}}(2019)}]{HAWC-ICRC2019-176}
\bibinfo{author}{\bibfnamefont{J.~C.} \bibnamefont{{Arteaga-Velazquez}}}
  \bibnamefont{and} \bibinfo{author}{\bibfnamefont{J.~D.}
  \bibnamefont{{Alvarez}}}, \bibinfo{journal}{Proceedings of Science}
  \textbf{\bibinfo{volume}{ICRC2019}}, \bibinfo{pages}{176}
  (\bibinfo{year}{2019}).

\bibitem[{\citenamefont{{Apel} et~al.}(2013)\citenamefont{{Apel},
  {Arteaga-Vel{\'a}zquez}, {Bekk}, {Bertaina}, {Bl{\"u}mer}, {Bozdog},
  {Brancus}, {Cantoni}, {Chiavassa}, {Cossavella}
  et~al.}}]{2013APh....47...54A}
\bibinfo{author}{\bibfnamefont{W.~D.} \bibnamefont{{Apel}}},
  \bibnamefont{et~al.}, \bibinfo{journal}{Astroparticle Physics}
  \textbf{\bibinfo{volume}{47}}, \bibinfo{pages}{54} (\bibinfo{year}{2013}).

\bibitem[{\citenamefont{{Bartoli}
  et~al.}(2015{\natexlab{b}})\citenamefont{{Bartoli}, {Bernardini}, {Bi},
  {Cao}, {Catalanotti}, {Chen}, {Chen}, {Cui}, {Dai}, {D'Amone}
  et~al.}}]{2015PhRvD..91k2017B}
\bibinfo{author}{\bibfnamefont{B.}~\bibnamefont{{Bartoli}}},
  \bibnamefont{et~al.}, \bibinfo{journal}{\prd} \textbf{\bibinfo{volume}{91}},
  \bibinfo{eid}{112017} (\bibinfo{year}{2015}{\natexlab{b}}),
  \eprint{1503.07136}.

\bibitem[{\citenamefont{{H{\"o}randel}}(2003)}]{2003APh....19..193H}
\bibinfo{author}{\bibfnamefont{J.~R.} \bibnamefont{{H{\"o}randel}}},
  \bibinfo{journal}{Astropart. Phys.} \textbf{\bibinfo{volume}{19}},
  \bibinfo{pages}{193} (\bibinfo{year}{2003}), \eprint{astro-ph/0210453}.

\bibitem[{\citenamefont{{Zatsepin} and
  {Sokolskaya}}(2006)}]{2006A&A...458....1Z}
\bibinfo{author}{\bibfnamefont{V.~I.} \bibnamefont{{Zatsepin}}}
  \bibnamefont{and} \bibinfo{author}{\bibfnamefont{N.~V.}
  \bibnamefont{{Sokolskaya}}}, \bibinfo{journal}{\aap}
  \textbf{\bibinfo{volume}{458}}, \bibinfo{pages}{1} (\bibinfo{year}{2006}),
  \eprint{astro-ph/0601475}.

\bibitem[{\citenamefont{{Hillas}}(2006)}]{2006astro.ph..7109H}
\bibinfo{author}{\bibfnamefont{A.~M.} \bibnamefont{{Hillas}}},
  \bibinfo{journal}{arXiv Astrophysics e-prints}  (\bibinfo{year}{2006}),
  \eprint{astro-ph/0607109}.

\bibitem[{\citenamefont{{Gaisser}}(2012)}]{2012APh....35..801G}
\bibinfo{author}{\bibfnamefont{T.~K.} \bibnamefont{{Gaisser}}},
  \bibinfo{journal}{Astroparticle Physics} \textbf{\bibinfo{volume}{35}},
  \bibinfo{pages}{801} (\bibinfo{year}{2012}), \eprint{1111.6675}.

\bibitem[{\citenamefont{{Gaisser} et~al.}(2013)\citenamefont{{Gaisser},
  {Stanev}, and {Tilav}}}]{2013FrPhy...8..748G}
\bibinfo{author}{\bibfnamefont{T.~K.} \bibnamefont{{Gaisser}}},
  \bibinfo{author}{\bibfnamefont{T.}~\bibnamefont{{Stanev}}}, \bibnamefont{and}
  \bibinfo{author}{\bibfnamefont{S.}~\bibnamefont{{Tilav}}},
  \bibinfo{journal}{Frontiers of Physics} \textbf{\bibinfo{volume}{8}},
  \bibinfo{pages}{748} (\bibinfo{year}{2013}), \eprint{1303.3565}.

\bibitem[{\citenamefont{{Thoudam} et~al.}(2016)\citenamefont{{Thoudam},
  {Rachen}, {van Vliet}, {Achterberg}, {Buitink}, {Falcke}, and
  {H{\"o}randel}}}]{2016A&A...595A..33T}
\bibinfo{author}{\bibfnamefont{S.}~\bibnamefont{{Thoudam}}},
  \bibinfo{author}{\bibfnamefont{J.~P.} \bibnamefont{{Rachen}}},
  \bibinfo{author}{\bibfnamefont{A.}~\bibnamefont{{van Vliet}}},
  \bibinfo{author}{\bibfnamefont{A.}~\bibnamefont{{Achterberg}}},
  \bibinfo{author}{\bibfnamefont{S.}~\bibnamefont{{Buitink}}},
  \bibinfo{author}{\bibfnamefont{H.}~\bibnamefont{{Falcke}}}, \bibnamefont{and}
  \bibinfo{author}{\bibfnamefont{J.~R.} \bibnamefont{{H{\"o}randel}}},
  \bibinfo{journal}{\aap} \textbf{\bibinfo{volume}{595}}, \bibinfo{eid}{A33}
  (\bibinfo{year}{2016}), \eprint{1605.03111}.

\bibitem[{\citenamefont{{Guo} and
  {Yuan}}(2018{\natexlab{b}})}]{2018ChPhC..42g5103G}
\bibinfo{author}{\bibfnamefont{Y.-Q.} \bibnamefont{{Guo}}} \bibnamefont{and}
  \bibinfo{author}{\bibfnamefont{Q.}~\bibnamefont{{Yuan}}},
  \bibinfo{journal}{Chinese Physics C} \textbf{\bibinfo{volume}{42}},
  \bibinfo{eid}{075103} (\bibinfo{year}{2018}{\natexlab{b}}),
  \eprint{1701.07136}.

\bibitem[{\citenamefont{{Erlykin} and
  {Wolfendale}}(1997)}]{1997JPhG...23..979E}
\bibinfo{author}{\bibfnamefont{A.~D.} \bibnamefont{{Erlykin}}}
  \bibnamefont{and} \bibinfo{author}{\bibfnamefont{A.~W.}
  \bibnamefont{{Wolfendale}}}, \bibinfo{journal}{J. Phys. G Nucl. Phys.}
  \textbf{\bibinfo{volume}{23}}, \bibinfo{pages}{979} (\bibinfo{year}{1997}).

\bibitem[{\citenamefont{{Sveshnikova} et~al.}(2013)\citenamefont{{Sveshnikova},
  {Strelnikova}, and {Ptuskin}}}]{2013APh....50...33S}
\bibinfo{author}{\bibfnamefont{L.~G.} \bibnamefont{{Sveshnikova}}},
  \bibinfo{author}{\bibfnamefont{O.~N.} \bibnamefont{{Strelnikova}}},
  \bibnamefont{and} \bibinfo{author}{\bibfnamefont{V.~S.}
  \bibnamefont{{Ptuskin}}}, \bibinfo{journal}{Astroparticle Physics}
  \textbf{\bibinfo{volume}{50}}, \bibinfo{pages}{33} (\bibinfo{year}{2013}),
  \eprint{1301.2028}.

\bibitem[{\citenamefont{{Savchenko} et~al.}(2015)\citenamefont{{Savchenko},
  {Kachelrie{\ss}}, and {Semikoz}}}]{2015ApJ...809L..23S}
\bibinfo{author}{\bibfnamefont{V.}~\bibnamefont{{Savchenko}}},
  \bibinfo{author}{\bibfnamefont{M.}~\bibnamefont{{Kachelrie{\ss}}}},
  \bibnamefont{and} \bibinfo{author}{\bibfnamefont{D.~V.}
  \bibnamefont{{Semikoz}}}, \bibinfo{journal}{\apjl}
  \textbf{\bibinfo{volume}{809}}, \bibinfo{eid}{L23} (\bibinfo{year}{2015}),
  \eprint{1505.02720}.

\bibitem[{\citenamefont{{Liu} et~al.}(2019)\citenamefont{{Liu}, {Guo}, and
  {Yuan}}}]{2019JCAP...10..010L}
\bibinfo{author}{\bibfnamefont{W.}~\bibnamefont{{Liu}}},
  \bibinfo{author}{\bibfnamefont{Y.-Q.} \bibnamefont{{Guo}}}, \bibnamefont{and}
  \bibinfo{author}{\bibfnamefont{Q.}~\bibnamefont{{Yuan}}},
  \bibinfo{journal}{\jcap} \textbf{\bibinfo{volume}{10}}, \bibinfo{eid}{010}
  (\bibinfo{year}{2019}), \eprint{1812.09673}.

\bibitem[{\citenamefont{{Qu}}(2019)}]{2019arXiv190100249Q}
\bibinfo{author}{\bibfnamefont{X.}~\bibnamefont{{Qu}}}, \bibinfo{journal}{arXiv
  e-prints}  (\bibinfo{year}{2019}), \eprint{1901.00249}.

\bibitem[{\citenamefont{{Qiao} et~al.}(2019)\citenamefont{{Qiao}, {Liu}, {Guo},
  and {Yuan}}}]{2019arXiv190512505Q}
\bibinfo{author}{\bibfnamefont{B.-Q.} \bibnamefont{{Qiao}}},
  \bibinfo{author}{\bibfnamefont{W.}~\bibnamefont{{Liu}}},
  \bibinfo{author}{\bibfnamefont{Y.-Q.} \bibnamefont{{Guo}}}, \bibnamefont{and}
  \bibinfo{author}{\bibfnamefont{Q.}~\bibnamefont{{Yuan}}},
  \bibinfo{journal}{arXiv e-prints}  (\bibinfo{year}{2019}),
  \eprint{1905.12505}.

\bibitem[{\citenamefont{{Karmanov} et~al.}(2019)\citenamefont{{Karmanov},
  {Kovalev}, {Kudryashov}, {Kurganov}, {Latonov}, {Panov}, {Podorozhnyy}, and
  {Turundaevskiy}}}]{2019arXiv190705987K}
\bibinfo{author}{\bibfnamefont{D.}~\bibnamefont{{Karmanov}}},
  \bibinfo{author}{\bibfnamefont{I.}~\bibnamefont{{Kovalev}}},
  \bibinfo{author}{\bibfnamefont{I.}~\bibnamefont{{Kudryashov}}},
  \bibinfo{author}{\bibfnamefont{A.}~\bibnamefont{{Kurganov}}},
  \bibinfo{author}{\bibfnamefont{V.}~\bibnamefont{{Latonov}}},
  \bibinfo{author}{\bibfnamefont{A.}~\bibnamefont{{Panov}}},
  \bibinfo{author}{\bibfnamefont{D.}~\bibnamefont{{Podorozhnyy}}},
  \bibnamefont{and}
  \bibinfo{author}{\bibfnamefont{A.}~\bibnamefont{{Turundaevskiy}}},
  \bibinfo{journal}{arXiv e-prints} \bibinfo{eid}{arXiv:1907.05987}
  (\bibinfo{year}{2019}), \eprint{1907.05987}.

\bibitem[{\citenamefont{{Yoon} et~al.}(2011)\citenamefont{{Yoon}, {Ahn},
  {Allison}, {Bagliesi}, {Beatty}, {Bigongiari}, {Boyle}, {Childers},
  {Conklin}, {Coutu} et~al.}}]{2011ApJ...728..122Y}
\bibinfo{author}{\bibfnamefont{Y.~S.} \bibnamefont{{Yoon}}},
  \bibnamefont{et~al.}, \bibinfo{journal}{\apj} \textbf{\bibinfo{volume}{728}},
  \bibinfo{eid}{122} (\bibinfo{year}{2011}), \eprint{1102.2575}.

\bibitem[{\citenamefont{{Lipari} and {Vernetto}}(2019)}]{2019arXiv191101311L}
\bibinfo{author}{\bibfnamefont{P.}~\bibnamefont{{Lipari}}} \bibnamefont{and}
  \bibinfo{author}{\bibfnamefont{S.}~\bibnamefont{{Vernetto}}},
  \bibinfo{journal}{arXiv e-prints} \bibinfo{eid}{arXiv:1911.01311}
  (\bibinfo{year}{2019}), \eprint{1911.01311}.

\bibitem[{\citenamefont{{Ahn} et~al.}(2009)\citenamefont{{Ahn}, {Allison},
  {Bagliesi}, {Barbier}, {Beatty}, {Bigongiari}, {Brandt}, {Childers},
  {Conklin}, {Coutu} et~al.}}]{2009ApJ...707..593A}
\bibinfo{author}{\bibfnamefont{H.~S.} \bibnamefont{{Ahn}}},
  \bibnamefont{et~al.}, \bibinfo{journal}{\apj} \textbf{\bibinfo{volume}{707}},
  \bibinfo{pages}{593} (\bibinfo{year}{2009}), \eprint{0911.1889}.

\bibitem[{\citenamefont{{Aglietta} et~al.}(1996)\citenamefont{{Aglietta},
  {Alessandro}, {Antonioli}, {Arneodo}, {Bergamasco}, {Bertaina}, {Bosio},
  {Castellina}, {Castagnoli}, {Chiavassa} et~al.}}]{1996ApJ...470..501A}
\bibinfo{author}{\bibfnamefont{M.}~\bibnamefont{{Aglietta}}},
  \bibnamefont{et~al.}, \bibinfo{journal}{\apj} \textbf{\bibinfo{volume}{470}},
  \bibinfo{pages}{501} (\bibinfo{year}{1996}).

\bibitem[{\citenamefont{{Amenomori} et~al.}(2006)\citenamefont{{Amenomori},
  {Ayabe}, {Bi}, {Chen}, {Cui}, {Danzengluobu}, {Ding}, {Ding}, {Feng}, {Feng}
  et~al.}}]{2006Sci...314..439A}
\bibinfo{author}{\bibfnamefont{M.}~\bibnamefont{{Amenomori}}},
  \bibnamefont{et~al.}, \bibinfo{journal}{Science}
  \textbf{\bibinfo{volume}{314}}, \bibinfo{pages}{439} (\bibinfo{year}{2006}),
  \eprint{astro-ph/0610671}.

\bibitem[{\citenamefont{{Aglietta} et~al.}(2009)\citenamefont{{Aglietta},
  {Alekseenko}, {Alessandro}, {Antonioli}, {Arneodo}, {Bergamasco}, {Bertaina},
  {Bonino}, {Castellina}, {Chiavassa} et~al.}}]{2009ApJ...692L.130A}
\bibinfo{author}{\bibfnamefont{M.}~\bibnamefont{{Aglietta}}},
  \bibnamefont{et~al.}, \bibinfo{journal}{\apjl}
  \textbf{\bibinfo{volume}{692}}, \bibinfo{pages}{L130} (\bibinfo{year}{2009}),
  \eprint{0901.2740}.

\bibitem[{\citenamefont{{Aartsen} et~al.}(2016)\citenamefont{{Aartsen},
  {Abraham}, {Ackermann}, {Adams}, {Aguilar}, {Ahlers}, {Ahrens}, {Altmann},
  {Anderson}, {Ansseau} et~al.}}]{2016ApJ...826..220A}
\bibinfo{author}{\bibfnamefont{M.~G.} \bibnamefont{{Aartsen}}},
  \bibnamefont{et~al.}, \bibinfo{journal}{\apj} \textbf{\bibinfo{volume}{826}},
  \bibinfo{eid}{220} (\bibinfo{year}{2016}), \eprint{1603.01227}.

\bibitem[{\citenamefont{{Amenomori} et~al.}(2017)\citenamefont{{Amenomori},
  {Bi}, {Chen}, {Chen}, {Chen}, {Cui}, {Danzengluobu}, {Ding}, {Feng}, {Feng}
  et~al.}}]{2017ApJ...836..153A}
\bibinfo{author}{\bibfnamefont{M.}~\bibnamefont{{Amenomori}}},
  \bibnamefont{et~al.}, \bibinfo{journal}{\apj} \textbf{\bibinfo{volume}{836}},
  \bibinfo{eid}{153} (\bibinfo{year}{2017}), \eprint{1701.07144}.

\bibitem[{\citenamefont{{Bai} et~al.}(2019)\citenamefont{{Bai}, {Bi}, {Bi},
  {Cao}, {Chen}, {Chen}, {Chiavassa}, {Cui}, {Dai}, {della Volpe}
  et~al.}}]{2019arXiv190502773B}
\bibinfo{author}{\bibfnamefont{X.}~\bibnamefont{{Bai}}}, \bibnamefont{et~al.},
  \bibinfo{journal}{arXiv e-prints}  (\bibinfo{year}{2019}),
  \eprint{1905.02773}.

\bibitem[{\citenamefont{{Zhang} et~al.}(2014)\citenamefont{{Zhang}, {Adriani},
  {Albergo}, {Ambrosi}, {An}, {Bao}, {Battiston}, {Bi}, {Cao}, {Chai}
  et~al.}}]{2014SPIE.9144E..0XZ}
\bibinfo{author}{\bibfnamefont{S.~N.} \bibnamefont{{Zhang}}},
  \bibnamefont{et~al.}, in \emph{\bibinfo{booktitle}{Proc. SPIE}}
  (\bibinfo{year}{2014}), vol. \bibinfo{volume}{9144}, p.
  \bibinfo{pages}{91440X}, \eprint{1407.4866}.

\end{thebibliography}

\end{document}